\documentclass[%
  reprint,
  superscriptaddress,
  amsmath,
  amssymb,
  aps,
  floatfix,
  showkeys
]{revtex4-2}

\usepackage{graphicx}
\usepackage{dcolumn}
\usepackage{multirow}
\usepackage[dvipsnames]{xcolor}
\usepackage{float}
\usepackage{makecell}
\usepackage{verbatim}
\usepackage{xspace}
\usepackage{siunitx}
\usepackage[version=4]{mhchem}
\usepackage[colorlinks=true, linkcolor=NavyBlue, citecolor=NavyBlue, urlcolor=gray]{hyperref}

\graphicspath{{./}}

\newcommand{\sifull}[0]{Supplemental Material (SM)\xspace}
\newcommand{\sishort}[0]{SM\xspace}

\newcommand*{\fref}[1]{Fig.~\ref{#1}}
\newcommand*{\tref}[1]{Table~\ref{#1}}

\newcommand*{\sref}[1]{Section~\ref{#1}}

\newcommand*{\olcite}[1]{Ref.~\cite{#1}}
\newcommand*{\method}[1]{\hyperref[#1]{Methods}}
\newcommand*{\discussion}[1]{\hyperref[#1]{Discussion}}

\newcommand{\pnma}[0]{Pnma\xspace}
\newcommand{\pmtmo}[0]{P$\bar{3}$m1\xspace}
\newcommand{\lyc}[0]{\ce{Li3YCl6}\xspace}

\begin{document}

\title{Strain-Dependent Ionic Transport in \lyc Solid Electrolytes}

\author{Wei-Fan Huang}
\thanks{Equal contribution}
\author{Jin Dai}
\thanks{Equal contribution}
\affiliation{William A. Brookshire Department of Chemical and Biomolecular Engineering, University of Houston, Houston, TX, 77204, USA}
\author{Jiahui Pan}
\affiliation{Institute of Fundamental and Frontier Sciences, University of Electronic Science and Technology of China, Chengdu, 611731, China}
\author{Mingjian Wen}
\email{mjwen@uestc.edu.cn}
\affiliation{Institute of Fundamental and Frontier Sciences, University of Electronic Science and Technology of China, Chengdu, 611731, China}
\affiliation{Frontier Center of Energy Distribution and Integration, Tianfu Jiangxi Laboratory, Chengdu, 641419, China}

\date{\today}

\begin{abstract}
  Solid-state batteries require electrolytes that sustain high ionic conductivity under the mechanical environment of a functioning cell.
  Lattice strain, arising from stack pressure, lattice mismatch at interfaces, and more importantly electrochemical cycling, can either enhance or suppress \ce{Li+} transport in solid electrolytes, yet how it couples to the underlying diffusion mechanism in \lyc remains poorly understood.
  Using \lyc halide superionic conductor, we address this with large-scale molecular dynamics simulations driven by an Atomic Cluster Expansion (ACE) machine learning interatomic potential trained on first-principles data.
  The ACE model faithfully reproduces experimental and \textit{ab initio} structural, mechanical, and transport properties of \lyc.
  Extending the simulations to high temperature reveals a crossover from predominantly one-dimensional hopping at low temperature toward diffusion through a connected three-dimensional pathway network at high temperature.
  Strain substantially modulates diffusivity: tensile strain enhances it while compressive strain suppresses it.
  Free-breakpoint model selection resolves two Arrhenius regimes for most strain states, and the fitted $T_c$ varies nonmonotonically with strain.
  Strain modifies the effective activation energies and the accessibility of diffusion pathways in a temperature-dependent manner.
  These results quantify the strain sensitivity of ionic conductivity and diffusion pathways in \lyc solid-state electrolytes.
\end{abstract}

\maketitle

\section{Introduction}
\label{sec:introduction}

Solid-state batteries (SSBs) are widely regarded as a promising successor to current lithium-ion technology, offering a pathway toward inherently safe and high-energy-density storage~\cite{Janek2016,Manthiram2017}.
Among the various technical requirements for high-performance SSBs, the ionic conductivity of the solid electrolyte (SE) stands as one of the most critical factors~\cite{Famprikis2019}.
High ionic conductivity underpins the power density and fast-charging capability required in practice~\cite{Bachman2016,Albertus2021,Kraft2017}.
Consequently, extensive research has focused on discovering and optimizing superionic conductors that can facilitate rapid transport at room temperature~\cite{Xiao2019,yang2022ionic}.

While high intrinsic ionic conductivity is a primary goal, the performance of these materials is often sensitive to the conditions encountered during cell integration.
A significant challenge stems from the poor interfacial contact between the rigid SE and the electrode materials, which can lead to high interfacial resistance~\cite{richards2016interface,Koerver2018,Xiao2019,Banerjee2020}.
To ensure mechanical integrity and continuous transport pathways, the application of external stack pressure is a common requirement in battery assembly~\cite{xu2024pressure,hu2024external,li2025critical,naik2025interrogating}.
Beyond stack pressure, lattice strain can also arise from lattice mismatch at the electrode--electrolyte interface~\cite{yang2021interfacial2,zhao2024unveiling} and, more importantly, from the electrode volume changes and interfacial reactions that accompany electrochemical cycling during battery operation~\cite{yu2019deformation,al2025chemo}.
Because the movement of lithium ions is closely linked to the local atomic environment, these induced strain fields can directly modify the ionic transport properties of the material~\cite{he2019interfacial, vzguns2022strain, li2025critical}.

Several atomistic studies have quantified how strain modifies \ce{Li+} transport in solid electrolytes~\cite{vzguns2022strain,moradabadi2020effect,lin2024strain,lin2024insight}.
In $\beta$-\ce{Li3PS4}, strain states that compress the $c$ axis raise the room-temperature conductivity by factors of 2 to 14, whereas those that stretch it lower it by comparable factors~\cite{vzguns2022strain}.
The response is not consistent from one material to the next, however: a compressive strain lowers the \ce{Li+} migration barrier in \ce{Al}-doped \ce{Li7La3Zr2O12} but raises it in \ce{Li10GeP2S12}~\cite{moradabadi2020effect}, and anti-perovskite \ce{Li3OCl} and \ce{Li3OBr} become more conductive under tension~\cite{lin2024strain,lin2024insight}.
This is expected if strain acts through the geometry of the diffusion channels and the rigidity of the anion framework, which are the features that govern ion transport in crystalline solid electrolytes~\cite{gao2020classical,song2024renewing,li2026ultrahigh} and which differ from one host to another.
Neither the sign nor the magnitude of the strain response therefore transfers between electrolytes, and the effect of a given deformation cannot be anticipated from its sense alone.

Two questions follow for any particular electrolyte.
The first is whether strain only changes how easily \ce{Li+} moves along the pathways that already carry the current, or whether it changes which pathways carry it.
The second is whether a temperature-dependent change of transport mechanism, where one exists, survives the deformation.
Both are sharply posed in \lyc, because prior simulations predict a crossover from one-dimensional hopping toward diffusion through a connected three-dimensional pathway network~\cite{schlem2021insights,qi2021bridging}, so strain can act on the connectivity of that network as well as on the individual barriers.
Answering them requires resolving the diffusion pathways themselves, rather than the aggregate conductivity alone.

We address these questions for \lyc, a halide electrolyte that combines high ionic conductivity with favorable mechanical deformability~\cite{asano2018solid,li2020progress,wang2022prospects}.
To resolve the diffusion pathways directly, we develop a machine learning interatomic potential based on the Atomic Cluster Expansion (ACE) framework~\cite{drautz2019atomic,lysogorskiy2021performant}.
The ACE model is trained on density functional theory (DFT) data using the optB88-vdW functional, which was previously shown to improve the lattice parameters and the agreement of low-temperature ionic conductivity and activation energy with experiment compared with PBE~\cite{klimevs2011van,qi2021bridging}.
This approach enables us to perform large-scale molecular dynamics (MD) simulations of the effects of isotropic strain on diffusivity and activation energy across a wide temperature range.
Such simulations are beyond the reach of \emph{ab initio} molecular dynamics (AIMD), which is typically restricted to small systems at high temperature.

The ACE model reproduces DFT energies, forces, structures, and stresses and agrees with available low-temperature transport measurements, supporting its use for the strain-dependent simulations.
Using this framework, we demonstrate that the efficiency of transport is strongly strain-dependent, whereas the crossover itself is not resolved for every strain state and its fitted temperature varies nonmonotonically.
Specifically, tensile strain enhances diffusivity by making additional Li configurations and diffusion pathways accessible and lowers the effective activation energy at elevated temperatures.
In contrast, compressive strain suppresses pathway connectivity and increases the effective activation energy at high temperatures.
Strain therefore tunes the accessibility and efficiency of the intrinsic pathways rather than replacing them.
Taken together, these results quantify how isotropic strain changes the temperature-dependent diffusion pathways and ionic conductivity of \lyc.

\section{Materials and Methods}
\label{sec:methods}

\subsection{\lyc Structures}
\label{sec:lyc structures}

\begin{figure}[tbh!]
  \centering
  \includegraphics[width=0.5\textwidth]{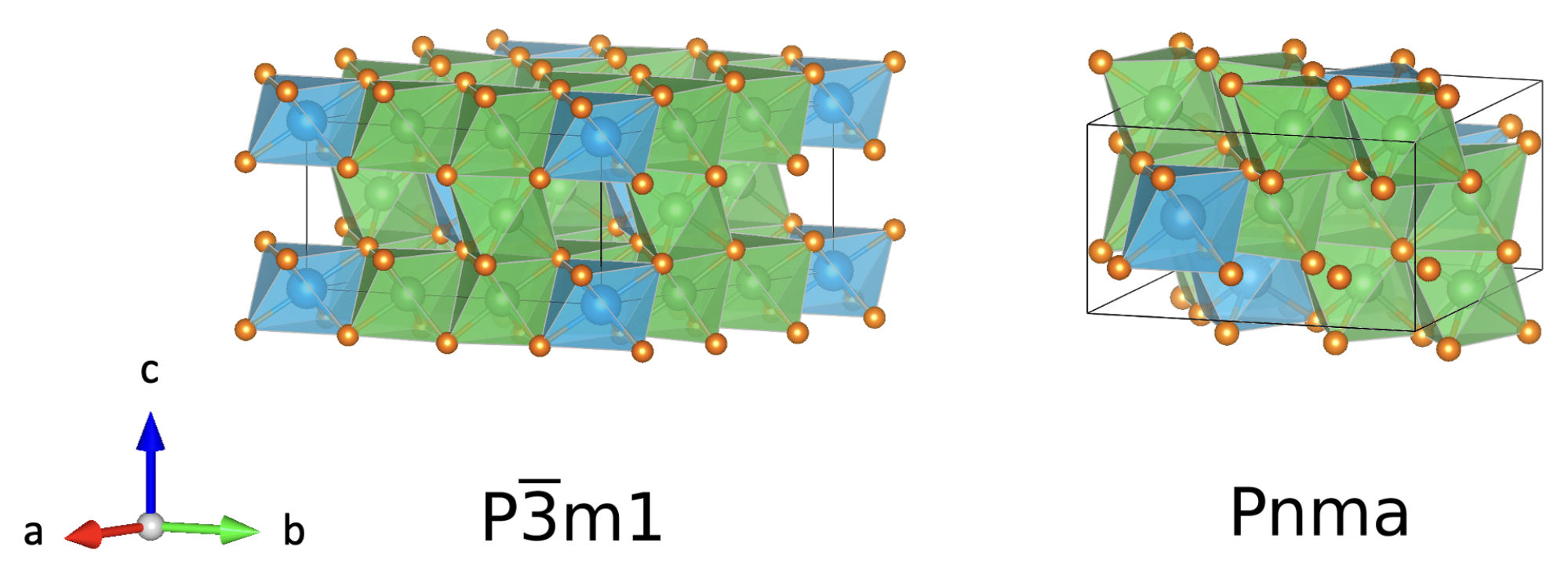}
  \caption{\textbf{Crystal structures of \lyc.}
    Cl, Li, and Y atoms are shown in orange, green, and blue, respectively.}
  \label{fig:structure}
\end{figure}

\lyc has various polymorphs~\cite{bohnsack1997ternare}, and this work focuses on the trigonal phase and the orthorhombic phase.
Both phases adopt a hexagonal close-packed anion sublattice of \ce{Cl-} anions, and they differ in the arrangements of \ce{Li+} and \ce{Y3+} cations within the \ce{Cl-} framework~\cite{liang2021metal}.
The trigonal phase has a space group of \pmtmo, which is isomorphic to \ce{Li3ErCl6} (ICSD No.\ 50151), whereas the orthorhombic phase has a space group of \pnma, which is an isomorph of \ce{Li3YbCl6} (ICSD No.\ 50152)~\cite{bohnsack1997ternare}.
\fref{fig:structure} depicts the crystal structures of both phases.
The \pmtmo phase consists of 30 atoms in the conventional cell, where Li occupy Wyckoff sites 6g and 6h with typical occupancies of 100\% and 50\%, respectively~\cite{schlem2021insights}.
For the atomistic models, we enumerated explicit orderings containing three Li atoms on the six 6h sites, which reduce by symmetry to three distinct configurations, of which the lowest-energy one was retained after DFT relaxation; the enumeration and selection procedure is described in Sec.~S1 of the \sifull.
The \pnma phase consists of 40 atoms in the conventional cell, and Li atoms reside on the 8d sites~\cite{liang2022series}.

\subsection{Atomic Cluster Expansion}

We adopt the Atomic Cluster Expansion (ACE) interatomic potential~\cite{drautz2019atomic,lysogorskiy2021performant} to model the \lyc system.
ACE provides a complete basis for representing atomic environments and thus interatomic potential energy surfaces.
It preserves fundamental physical symmetries, including translation, rotation, and inversion of space, as well as permutation of atomic species.

Within ACE, the total potential energy $E$ of a system consisting of $N$ atoms is constructed as a sum of the contributions of individual atoms:
\begin{equation}
  E = \sum_{i=1}^N \epsilon_i,
\end{equation}
where $\epsilon_i$ denotes the energy of atom $i$.
A description of the local environment of an atom is expressed as a linear combination of symmetry-adapted basis functions:
\begin{equation}
  \varphi_i = \sum_{\nu} c_{\nu} B_{i\nu},
  \label{eq:prop_expansion}
\end{equation}
where $B_{i\nu}$ are basis functions constructed from radial and angular parts, which satisfy the symmetry requirements.
The coefficients $c_{\nu}$ are learnable parameters.
For many-body interactions, the atomic energy is formulated as a nonlinear function $\mathcal{F}$ of one or more such properties:
\begin{equation}
  \epsilon_i = \mathcal{F}(\varphi_i^{1}, \varphi_i^{2}, \ldots, \varphi_i^{P}),
  \label{eq:energy_function}
\end{equation}
where $P$ is an index.
A common strategy uses a form inspired by the embedded-atom method (EAM)~\cite{daw1984embedded,daw1993embedded} and retains only the first two terms:
\begin{equation}
  \epsilon_i = \varphi_i^{1} + \sqrt{\varphi_i^{2}}.
  \label{eq:embedded_example}
\end{equation}

By systematically increasing the complexity of the basis set, the ACE framework provides controlled convergence toward the true potential energy surface, offering a robust and physically grounded representation of interatomic interactions.
The ACE framework enables the construction of systematically improvable, highly accurate, and efficient machine learning interatomic potentials for large-scale atomistic simulations.
A detailed discussion of the ACE framework is provided in Ref.~\cite{dusson2022atomic}.

\subsection{Development of ACE Potential}
\label{sec:ace:development}

\begin{figure*}[tbh!]
  \centering
  \includegraphics[width=0.8\linewidth]{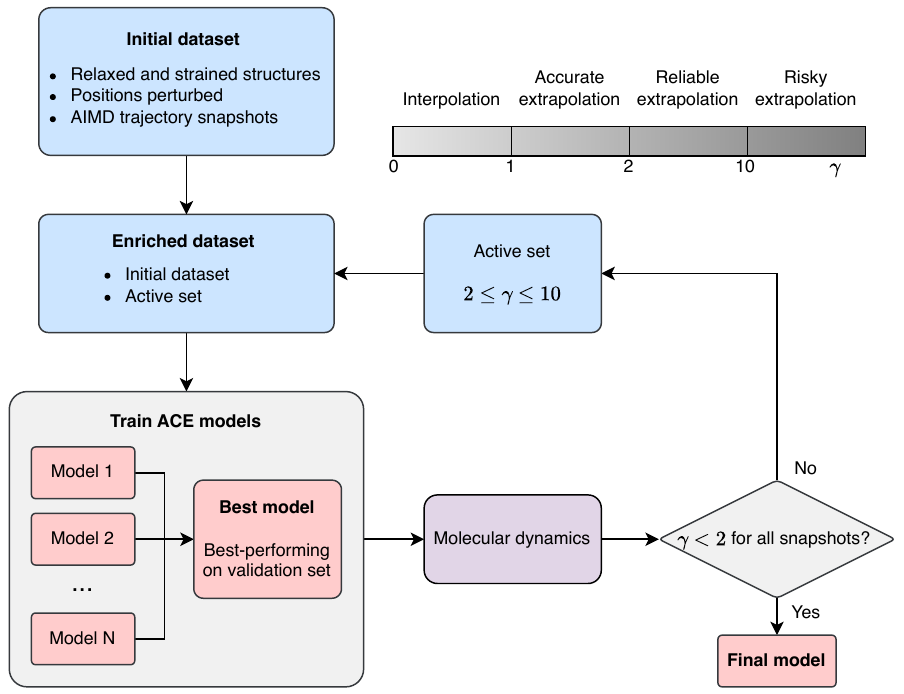}
  \caption{\textbf{Active learning workflow to train ACE models for \lyc.}
    With an initial dataset, the workflow iteratively improves model quality by adding new data for training.
    The loop stops when the extrapolation grade $\gamma$ for all molecular dynamics snapshots is smaller than a threshold of 2.
  }
  \label{fig:workflow}
\end{figure*}

The ACE model was trained on DFT reference energies and forces using an active learning strategy for multiple rounds (\fref{fig:workflow}).
An initial dataset of \lyc structures in both \pmtmo and \pnma space groups was generated to train the first generation of models.
The initial dataset consisted of three types of configurations:
(1) relaxed structures and structures under strains of $\pm 1$ and $\pm 2$\%;
(2) the relaxed and strained structures with perturbed atom positions, where the perturbation is drawn from a uniform distribution in the range 0 to 0.3~\AA;
and
(3) structures sampled from AIMD trajectories at 500~K.
This led to an initial dataset of 2115 structures, whose total energy and atomic forces were obtained from DFT calculations.

The static DFT reference energies and forces used to train the production ACE model were calculated using the Vienna Ab initio Simulation Package (VASP)~\cite{kresse1996efficient} and orchestrated by the \texttt{atomate2}~\cite{ganose2025atomate2} workflow engine.
The calculations used the projector augmented-wave method and the optB88-vdW exchange--correlation functional, 60-atom \pmtmo and 80-atom \pnma cells, each a $1\times1\times2$ supercell of the corresponding conventional cell, a 520~eV plane-wave cutoff, and a $2\times2\times2$ $\Gamma$-centered $k$-point mesh.
Brillouin-zone integration used the tetrahedron method with Bl\"ochl corrections~\cite{blochl1994improved}.
The electronic self-consistency cycle was converged to $10^{-5}$~eV.
Structural relaxations allowed the ions, cell shape, and volume to relax, and were converged to residual forces below 0.02~eV/\AA, using the same functional, plane-wave cutoff, and $k$-point mesh as the static calculations.

The first-generation ACE model was trained on the initial dataset.
The dataset was randomly split into training, validation, and test subsets with a ratio of 8:1:1.
Model parameters were optimized using the training set by minimizing a loss function between the energy and forces predicted by ACE and the reference DFT values.
Hyperparameters were determined by model performance on the validation set, and results are reported on the test set.

The trained first-generation ACE model was then continuously improved using an active learning strategy.
This was achieved in three steps.
First, MD simulations using the trained ACE model were performed for a maximum of 500000 steps.
These exploration simulations used the NVT ensemble at 300, 500, and 700~K and sampled both relaxed structures and structures under strains of $\pm1$ and $\pm2$\%.
Second, for each snapshot (i.e., structure) along the MD trajectory, the \emph{extrapolation grade} $\gamma$ was calculated using the D-optimality criterion~\cite{podryabinkin2023mlip,novikov2020mlip}.
In this approach, the active set contains linearly independent local atomic environments selected from the training data.
Each environment in a new MD snapshot is compared with this active set, and the grade of the snapshot is the maximum grade among all its atomic environments; a larger $\gamma$ indicates a greater degree of extrapolation~\cite{podryabinkin2023mlip}.
Following \olcite{novikov2020mlip}, we employed two thresholds: $\gamma_{\text{select}}=2$ and $\gamma_{\text{break}}=10$.
Structures with $\gamma < \gamma_{\text{select}}$ fall within the interpolation or accurate extrapolation region and require no intervention.
A structure with $\gamma > \gamma_{\text{break}}$ is flagged as risky extrapolation.
Consequently, the MD simulation is terminated upon encountering such a structure, regardless of whether the 500000-step limit has been reached.
Finally, structures in the intermediate range, $\gamma_{\text{select}} \leq \gamma \leq \gamma_{\text{break}}$, are classified as reliable extrapolations and are selected to enrich the dataset.
The enriched dataset was then used to retrain the ACE model.
This active learning loop continued until all snapshots along the MD trajectory achieved a $\gamma$ smaller than $\gamma_\text{select}$.
After seven enrichment cycles, the final dataset contained 10,062 DFT-labeled structures, of which 6200 (61.62\%) were strained configurations.

At each iteration, the newly labeled configurations were split using the same 8:1:1 ratio and appended to the respective subsets, so the final held-out test set includes every enrichment cycle.
The snapshots generated during the active learning MD also served as an active test set, on which the extrapolation grade was evaluated to assess how well the training data cover the configuration space that the model is applied to.
Further ACE fitting details such as the hyperparameter search are provided in Sec.~S1 of the \sishort.

\section{Results and Discussion}
\label{sec:results:and:discussion}

\subsection{Validation of the Developed ACE Potential}
\label{sec:validation}

\begin{figure*}[tbh!]
  \centering
  \includegraphics[width=0.7\textwidth]{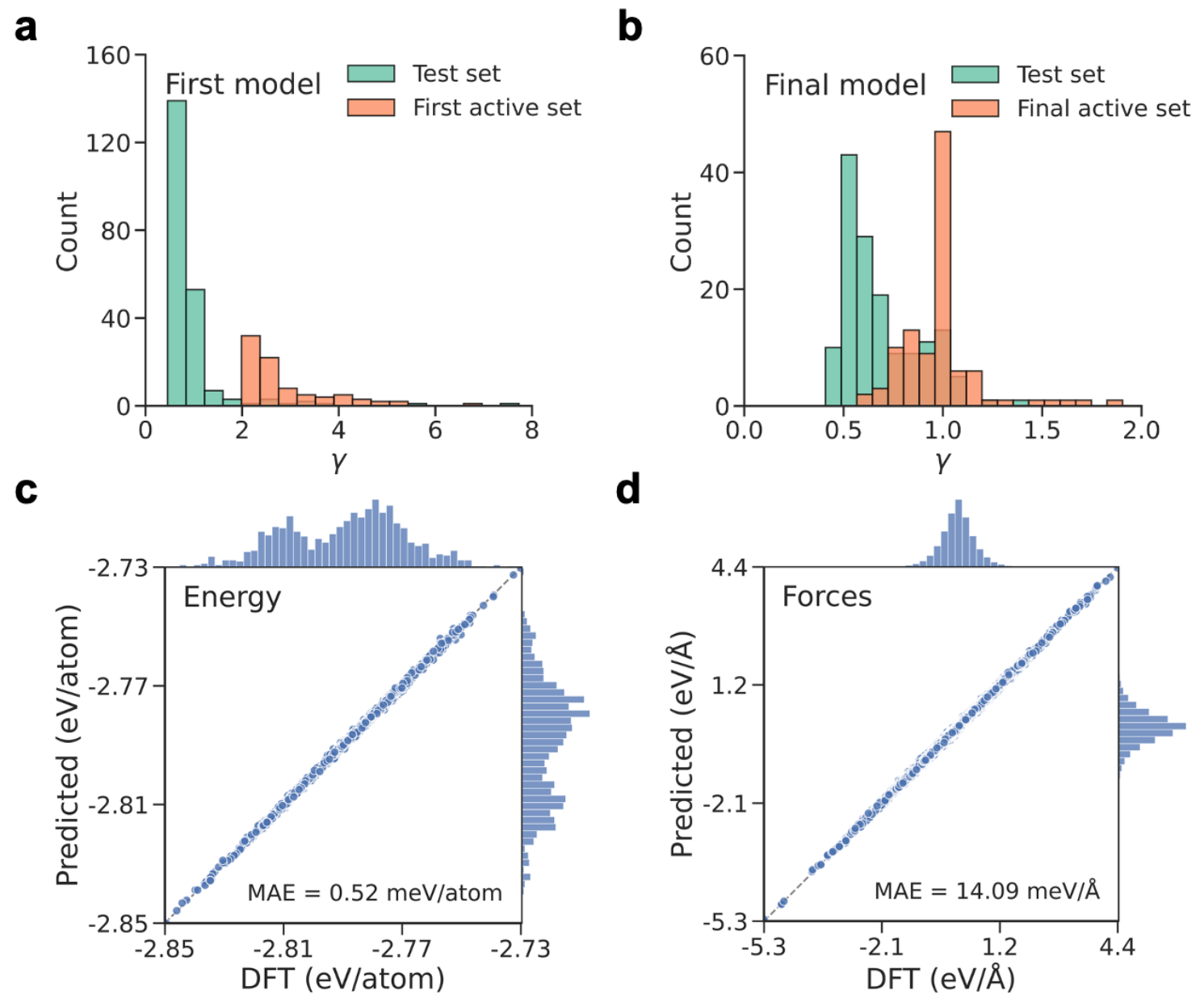}
  \caption{\textbf{Performance of the developed ACE model.}
    Histogram of the maximum extrapolation grade $\gamma_\text{max}$ in the active learning loop: (a) first model and (b) final model.
    ACE predicted versus DFT reference (c) energy and (d) atomic forces on the test set obtained using the final ACE model.
  }
  \label{fig:performance:testset}
\end{figure*}

\begin{table}[tbh!]
  \centering
  \caption{\textbf{Comparison of computed and experimental lattice parameters.}
    The angles between the lattice vectors are $\alpha=\beta=90^\circ$ and $\gamma=120^\circ$ for the \pmtmo phase, and $\alpha=\beta=\gamma=90^\circ$ for the \pnma phase.
  }
  \label{tab:lattice:params}
  \begin{tabular}{cc c c c c c c}
    \hline
     & $a$ (\AA) & $b$ (\AA) & $c$ (\AA) & Method                          \\
    \hline
    \multirow{3}{*}{\pmtmo}
     & 11.21     & 11.21     & 6.05      & Exp.~\cite{schlem2021insights}  \\
     & 11.03     & 11.03     & 6.09      & DFT (optB88-vdW), ours          \\
     & 11.06     & 11.04     & 6.05      & ACE, ours                       \\
    \hline
    \multirow{3}{*}{\pnma}
     & 12.93     & 11.12     & 6.04      & Exp.~\cite{bohnsack1997ternare} \\
     & 12.76     & 11.06     & 6.09      & DFT (optB88-vdW), ours          \\
     & 12.83     & 11.03     & 6.10      & ACE, ours                       \\
    \hline
  \end{tabular}
\end{table}

The active learning loop proceeded for seven enrichment cycles, producing eight generations of ACE models, and the final model was obtained using all data accumulated thus far.
\fref{fig:performance:testset}a and b show the extrapolation grade, $\gamma$, distributions for the first and final ACE models, respectively.
For the first ACE model, although most configurations from the test set of the initial dataset exhibit $\gamma$ values below 2, all configurations from the active test set have $\gamma$ values exceeding the selection threshold of 2.
This suggested that the model was significantly extrapolating and became unreliable in the MD simulations.
The model continued to be improved through active learning.
For the final ACE model, all configurations from both the test set and the active test set have $\gamma<2$, that is, they fall within the interpolation region and require no additional training data.
The active test set consists of snapshots taken from the active learning MD trajectories, which sampled temperatures of 300, 500, and 700~K and strains of $\pm1$ and $\pm2$\%.

The energy and forces predicted by the final ACE model closely match the DFT reference values, as shown in \fref{fig:performance:testset}c and d.
It achieves mean absolute errors (MAEs) of 0.52~meV/atom for energy and 14.09~meV/\AA\ for forces on the test set.
Energy and force MAEs for each generation of ACE models during active learning are summarized in Tables~S2 and~S3 in the \sishort.
We also checked the structural properties predicted by the ACE model.
The ACE lattice parameters were obtained by geometry optimization of both the cell and the atomic positions.
The predicted lattice parameters of both \pmtmo and \pnma phases closely match the DFT values, with deviations within 0.7\%, and lie within 1.6\% of experiment (\tref{tab:lattice:params}).

In addition, we tested the model's ability to predict mechanical properties by calculating the elastic tensor.
The elastic tensor provides a fundamental and complete description of a material's response to small deformations, capturing how stress and strain interrelate in different crystallographic directions~\cite{nye1985physical}.

\begin{figure}[thb!]
  \centering
  \includegraphics[width=0.4\textwidth]{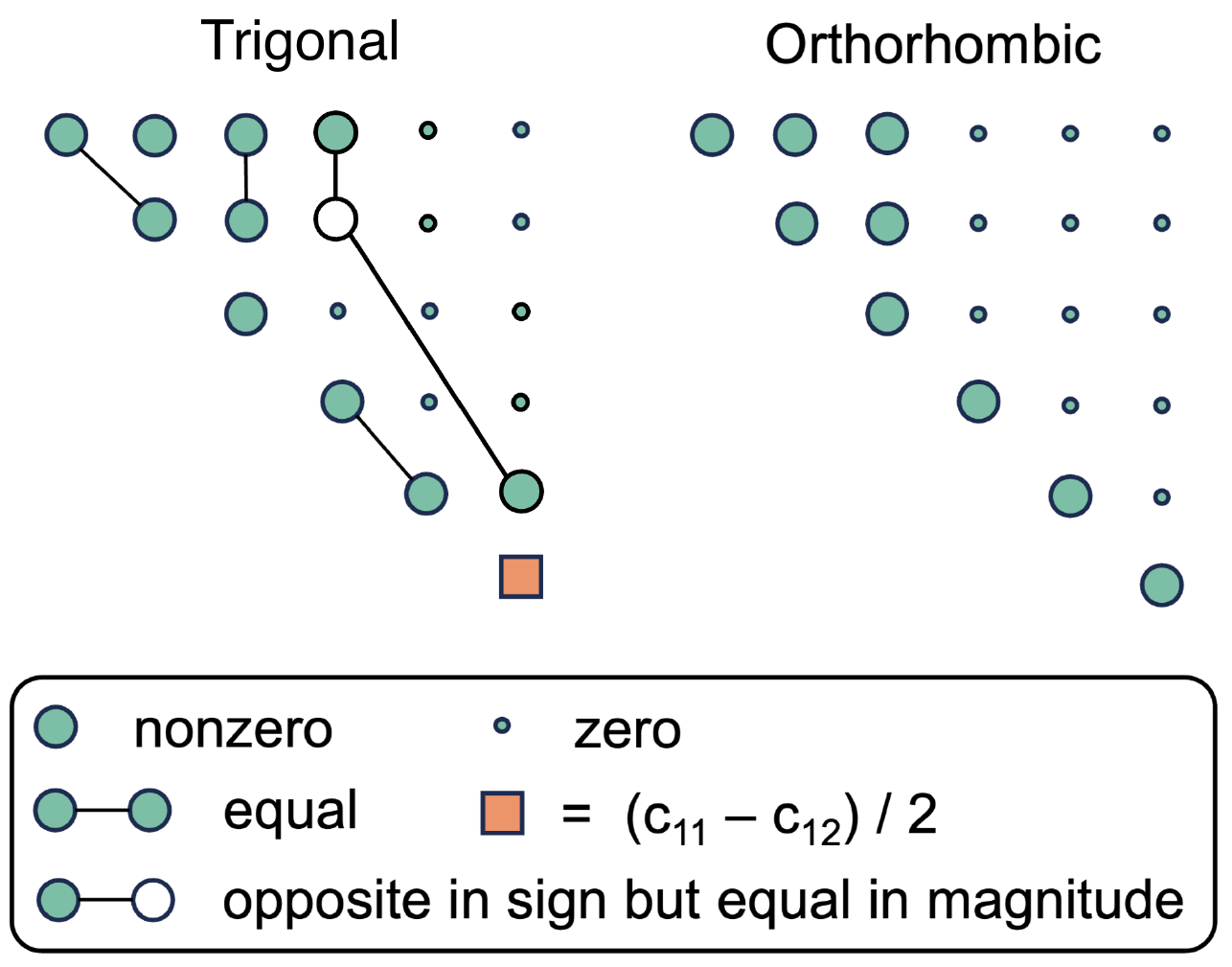}
  \caption{\textbf{Independent components of the elastic tensor of \lyc.}
    The trigonal \pmtmo phase and orthorhombic \pnma phase have six and nine independent components, respectively.
    Only the upper triangular part of the elastic tensor is shown, and the lower triangular part can be obtained by symmetry, i.e., $C_{ij} = C_{ji}$.
  }
  \label{fig:symmtery:class}
\end{figure}

The fourth-rank elastic tensor can be represented as a 6$\times$6 matrix using Voigt notation, where each component $C_{ij}$ relates stress to strain in specific crystallographic directions~\cite{nye1985physical}.
It has up to 21 independent components in the most general case, but for crystals with inherent symmetry, the number of independent components reduces significantly.
For \lyc, the \pmtmo phase that belongs to the trigonal crystal system has six independent components, whereas the \pnma phase that belongs to the orthorhombic crystal system has nine independent components~\cite{wen2024equivariant}.
\fref{fig:symmtery:class} illustrates the independent components and their symmetry-induced relationships.

\begin{table*}[tbh!]
  \centering
  \caption{\textbf{Computed elastic tensors and the derived elastic moduli.}
    The components of the elastic tensor $C_{ij}$ and the bulk modulus $B$, shear modulus $G$, and Young's modulus $E$ are reported in GPa.
    Only independent $C_{ij}$ components are listed; others can be obtained by symmetry as shown in \fref{fig:symmtery:class}.
    Every $B$, $G$, and $E$ in the table was computed in this work using the Hill averaging scheme; for the literature rows they were obtained from the published $C_{ij}$ rather than taken from the original papers.
  }
  \label{tab:elastic:tensors:moduli}
  \begin{tabular}{ccccccccccccccc}
    \hline
                            & $C_{11}$ & $C_{12}$ & $C_{13}$ & $C_{14}$ & $C_{22}$ & $C_{23}$ & $C_{33}$ & $C_{44}$ & $C_{55}$ & $C_{66}$ & $B$  & $G$  & $E$  & Method                              \\
    \hline
    \multirow{4}{*}{\pmtmo} & 50.5     & 18.5     & 6.8      & $-2.5$   &          &          & 40.8     & 12.6     &          &          & 22.5 & 15.0 & 36.9 & DFT (PBE+$U$)~\cite{qiu2021insights} \\
                            & 48.1     & 18.2     & 10.9     & $-0.1$   &          &          & 37.8     & 15.0     &          &          & 23.4 & 15.2 & 37.5 & DFT (PBE)~\cite{jiang2020materials} \\
                            & 54.4     & 19.9     & 13.6     & $-1.9$   &          &          & 43.2     & 13.2     &          &          & 27.0 & 15.4 & 38.8 & DFT (optB88-vdW), ours             \\
                            & 54.7     & 20.4     & 13.1     & $-1.7$   &          &          & 41.4     & 13.7     &          &          & 26.6 & 15.6 & 39.2 & ACE, ours                           \\
    \hline
    \multirow{2}{*}{\pnma}  & 52.4     & 18.7     & 12.6     &          & 51.5     & 11.8     & 39.4     & 12.2     & 11.4     & 16.4     & 25.1 & 14.3 & 36.2 & DFT (optB88-vdW), ours             \\
                            & 53.5     & 19.0     & 12.3     &          & 52.9     & 12.5     & 38.6     & 12.9     & 11.6     & 16.9     & 25.3 & 14.7 & 37.0 & ACE, ours                           \\
    \hline
  \end{tabular}
\end{table*}

We compute the elastic tensors with the finite difference method~\cite{ganose2025atomate2} using ACE and compare with DFT calculations.
The results are summarized in \tref{tab:elastic:tensors:moduli}.
The elastic tensor predicted by ACE closely matches those obtained from DFT using the same underlying optB88-vdW functional, with a maximum deviation of 10\% for $C_{14}$ in the \pmtmo phase ($-1.7$ vs $-1.9$~GPa).
For the other components, the deviation is even smaller (within 6\%), which represents good agreement given the complexity of the elastic tensor and the fact that it was not explicitly included in the training dataset.
In addition to the individual elastic tensor components, we also derive key mechanical moduli that characterize the material's response to various deformation modes.
Specifically, we computed the bulk modulus ($B$), shear modulus ($G$), and Young's modulus ($E$) using the Hill averaging scheme~\cite{nye1985physical} (computing procedures provided in Sec.~S5 of the \sishort).
Given that the individual components of the elastic tensor are accurately reproduced, as expected the derived moduli also exhibit strong agreement between ACE predictions and DFT calculations.
The $B$, $G$, and $E$ values of the two literature studies in \tref{tab:elastic:tensors:moduli} do not follow from their reported $C_{ij}$ under a common averaging scheme, so we recomputed them from the reported $C_{ij}$ using the same procedure.

Although stresses were not included in the fitting loss, the final ACE model reproduces the DFT stress tensor over the $-2$\% to $+2$\% strain range, with strain-resolved MAEs of 0.066--0.095~GPa (Sec.~S2 and Fig.~S3 in the \sishort).
We additionally tested a representative Oct--Oct migration pathway in \pmtmo at $-2$\%, 0\%, and $+2$\% strain by evaluating ACE-optimized pathway images with DFT.
The forward barriers agree within 4~meV, and both methods predict nearly identical barrier reductions under tensile strain (Sec.~S3 and Fig.~S7 in the \sishort).
This comparison validates the energies along the ACE pathway rather than an independently optimized DFT pathway.

Together, these validation tests show that the developed ACE model accurately reproduces the DFT energetic, structural, and elastic properties of \lyc and agrees with the available experimental lattice parameters.
In the following sections, we employ the validated ACE model to perform large-scale MD simulations to investigate the \ce{Li+} transport behavior of \lyc, which is critical for its performance as a solid-state electrolyte.

\subsection{\ce{Li+} Transport}
\label{sec:li:transport}

Ionic conductivity governs \ce{Li+} transport kinetics and dictates both the power capability and energy efficiency of solid-state batteries~\cite{yang2022ionic,Bachman2016}.
Consequently, numerous computational efforts using AIMD have been devoted to quantifying the ionic conductivity of \ce{Li3YCl6} at ambient temperatures~\cite{wang2019lithium,park2020theoretical,hu2023revealing,kim2021material}.
However, these studies are often compromised by three primary methodological limitations that undermine their accuracy.
First, room-temperature estimates largely rely on extrapolations from high-temperature simulations, a practice that can overlook transitions in diffusion mechanisms~\cite{geng2024elucidating,wang2023frustration}.
Second, many simulations are restricted to short trajectories of $\sim$100~ps and small supercells of $\sim$100 atoms; these scales are insufficient to statistically resolve long-range lithium-ion migration or capture representative bulk behavior~\cite{park2020theoretical,wang2019lithium,kim2021material,hu2023revealing}.
Third, several investigations were conducted at temperatures exceeding the material's 800~K melting point, potentially introducing liquid-like dynamics that do not reflect the physics of a solid-state conductor~\cite{asano2018solid,geng2024elucidating}.
Collectively, these constraints introduce significant uncertainty and may lead to inaccurate assessments of the material's performance.

To overcome the aforementioned limitations, we performed large-scale MD simulations using the developed ACE model.
The ACE model provides the necessary computational efficiency to simulate large systems and extended timescales while maintaining near-first-principles accuracy, effectively bridging the gap between accuracy and computational efficiency.
Specifically, we employed $3 \times 3 \times 6$ supercells, containing 1620 atoms for the \pmtmo phase and 2160 atoms for the \pnma phase.
Each state point was sampled using three independent 10~ns production trajectories.
The simulations were performed using the Large-scale Atomic/Molecular Massively Parallel Simulator (LAMMPS)~\cite{plimpton1995fast} under the NVT ensemble using a Nos\'{e}--Hoover thermostat with a 1~ps damping time and a timestep of 1~fs.
Configurations were saved every 1~ps, and the stored production trajectories were analyzed in full.
By performing these simulations across a wide temperature range (300--800~K), we eliminate the need for high-temperature extrapolation and ensure all data points remain below the material's melting point to maintain solid-state integrity.
The diffusivity $D$ was computed using the Einstein relation~\cite{einstein1905molekularkinetischen}, which links the mean squared displacement (MSD) of the lithium ions to the simulation time.
From the calculated diffusivity, the ionic conductivity $\sigma$ can be estimated using the Nernst--Einstein equation~\cite{nernst1888kinetik}.
We also compared tracer and collective diffusivities to assess ion--ion correlations.
Although these correlations change the transport magnitude, they do not alter the strain trend in either phase; see Sec.~S4 of the \sishort for details.

\begin{figure}[tbh!]
  \centering
  \includegraphics[width=0.8\columnwidth]{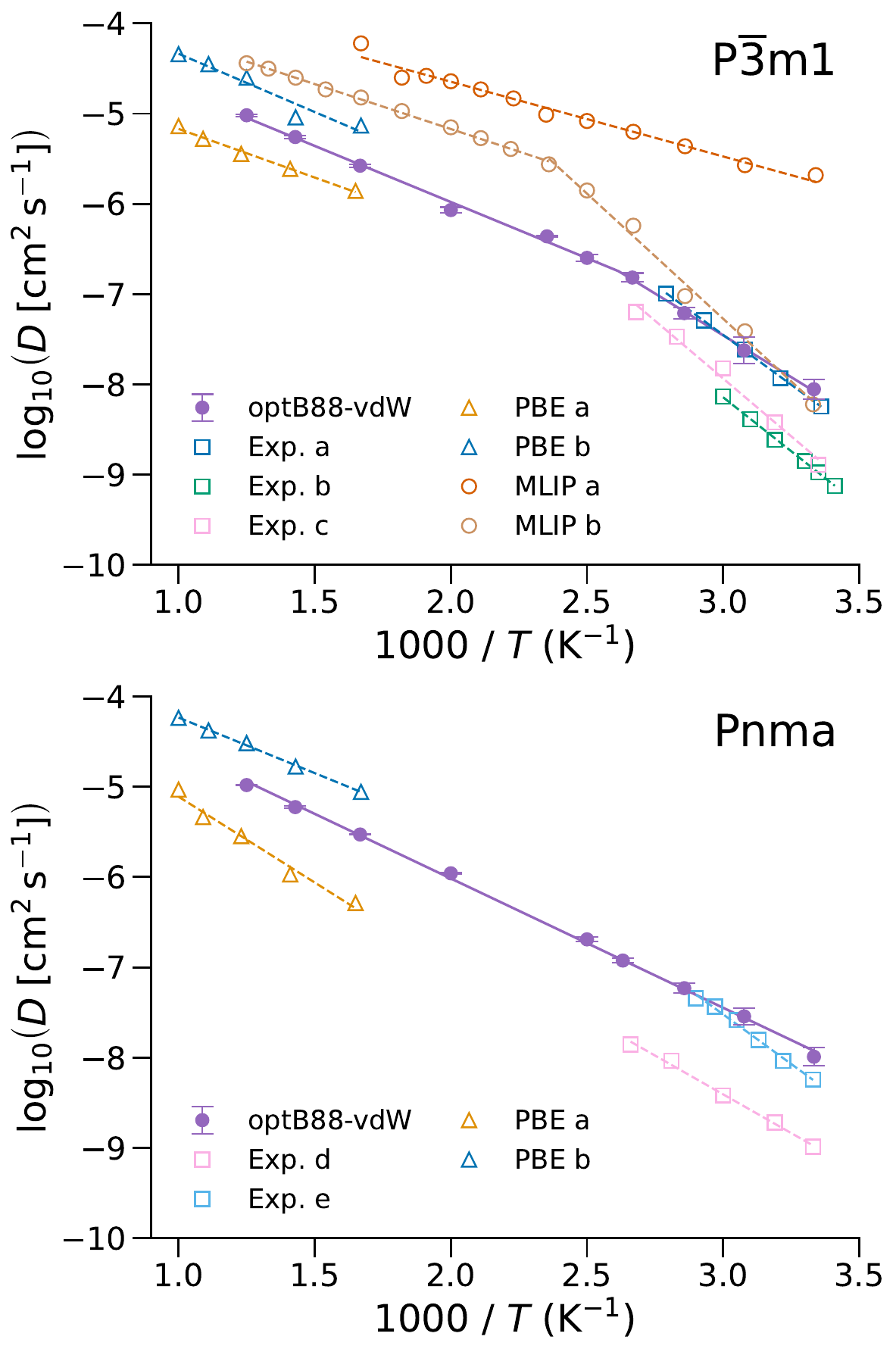}
  \caption{\textbf{Arrhenius plot of the \ce{Li+} diffusivity $D$}.
    Filled symbols represent results from this work, computed with the ACE model.
    Open symbols represent literature data.
    Experimental values: a~\cite{asano2018solid},  b~\cite{schlem2020mechanochemical}, c~\cite{chen2024accelerating}, d~\cite{hu2023revealing} and e~\cite{qiu2025scalable};
    Experimental values represent total polycrystalline-pellet conductivities rather than isolated bulk conductivities; measurement details are provided in the \sishort.
    AIMD simulations (PBE a~\cite{hu2023revealing} and PBE b~\cite{park2020theoretical});
    classical MD simulations using MLIPs (MLIP a, fitted to PBE DFT data)~\cite{qi2021bridging} and (MLIP b, fitted to  optB88-vdW DFT data)~\cite{qi2021bridging}.
    Error bars represent standard deviations from three independent MD runs.
  }
  \label{fig:sigma:vs:T}
\end{figure}

\begin{table}[tbh!]
  \centering
  \caption{\textbf{Comparison of the Li ionic conductivity $\sigma$ and activation energy $E_a$ in \lyc.}
    In the table, MD refers to classical MD simulation using interatomic potentials derived from DFT calculations.
    Rows labeled ``ours'' are Nernst--Einstein estimates derived from the Li diffusivity of our ACE model and are reported at 300~K; the corresponding values for our secondary PBEsol and PBEsol+MBD models are given in Table~S7 in the \sishort.
    Experimental values represent total conductivities of polycrystalline pellets.
    $\dagger$ indicates extrapolated values from high-temperature simulations.
  }
  \label{tab:sigma:Ea}
  \begin{tabular}{lcccc}
    \hline
    Phase  & $\sigma$        & $E_a$ & $T$       & Method                                     \\
           & (mS/cm)         & (eV)  & (K)       &                                            \\
    \hline
           & 0.51            & 0.40  & 240--360  & Exp.~\cite{asano2018solid}                 \\
           & 0.10            & 0.45  & 233--333  & Exp.~\cite{schlem2020mechanochemical}      \\
           & 0.11            & 0.51  & 300--373  & Exp.~\cite{chen2024accelerating}           \\
           & 12.17$^\dagger$ & 0.24  & 400--700  & AIMD (PBE)~\cite{chen2024accelerating}     \\
           & 14.00$^\dagger$ & 0.19  & 500--900  & AIMD (PBE)~\cite{wang2019lithium}          \\
           & 3.07$^\dagger$  & 0.15  & 600--1000 & AIMD (PBE)~\cite{hu2023revealing}          \\
           &                 & 0.26  & 600--1000 & AIMD (PBE)~\cite{park2020theoretical}      \\
           & 0.12$^\dagger$  & 0.37  & 700--1000 & AIMD (optB86b-vdW)~\cite{kim2021material}  \\
           & 152.66          & 0.16  & 300--800  & MD (PBE)~\cite{qi2021bridging}             \\
           & 0.56            & 0.49  & 300--425  & MD (optB88-vdW)~\cite{qi2021bridging}      \\
    \pmtmo &                 & 0.24  & 425--800  & MD (optB88-vdW)~\cite{qi2021bridging}      \\
           & 0.02$^\dagger$  & 0.80  & 375--425  & MD (optB88-vdW)~\cite{geng2024elucidating} \\
           &                 & 0.23  & 425--800  & MD (optB88-vdW)~\cite{geng2024elucidating} \\
           & 0.02$^\dagger$  & 0.74  & 330--425  & MD (optB88-vdW)~\cite{wang2023frustration} \\
           &                 & 0.24  & 425--700  & MD (optB88-vdW)~\cite{wang2023frustration} \\
           & 0.77            & 0.37  & 300--382  & MD (optB88-vdW), ours                       \\
           &                 & 0.25  & 382--800  & MD (optB88-vdW), ours                       \\
    \hline
           & 0.09            & 0.31  & 300--370  & Exp.~\cite{hu2023revealing}                \\
           & 0.50            & 0.36  & 300--348  & Exp.~\cite{qiu2025scalable}                \\
           & 0.04$^\dagger$  & 0.31  & 600--1000 & AIMD (PBE)~\cite{hu2023revealing}          \\
           &                 & 0.24  & 600--1000 & AIMD (PBE)~\cite{park2020theoretical}      \\
    \pnma  &                 & 0.69  & 300--380  & MD (optB88-vdW)~\cite{wang2023frustration} \\
           &                 & 0.21  & 380--700  & MD (optB88-vdW)~\cite{wang2023frustration} \\
           & 0.88            & 0.28  & 300--800  & MD (optB88-vdW), ours                       \\
    \hline
  \end{tabular}
\end{table}

To investigate the effect of temperature $T$ on \ce{Li+} diffusion, we plot the diffusivity $D$ as a function of $T$ for both the \pmtmo and \pnma phases in \fref{fig:sigma:vs:T}.
In the low-temperature range covered by experiment, the ACE model trained on optB88-vdW data reproduces the measured transport behavior closely, with a low-temperature $E_a$ of 0.37~eV for unstrained \pmtmo, in reasonable agreement with the experimental range of 0.40--0.51~eV~\cite{asano2018solid,schlem2020mechanochemical,chen2024accelerating}.
We also trained ACE models on PBEsol~\cite{perdew2008restoring} and on PBEsol+MBD~\cite{gould2016fractionally} reference data to test this choice; neither reproduces the measured low-temperature transport as closely, and adding the many-body dispersion treatment to PBEsol changes little (Fig.~S6 and Table~S7 in the \sishort).
This low-temperature agreement provides an experimental basis for the choice of reference data, and all simulations reported below use the ACE model trained on optB88-vdW data, referred to hereafter simply as the ACE model.

The temperature dependence is conventionally represented by the Arrhenius relationship~\cite{arrhenius1889reaktionsgeschwindigkeit}:
\begin{equation} \label{eq:arrhenius}
  D = D_0 \exp\left( -\frac{E_a}{k_B T} \right),
\end{equation}
where $D_0$ is a pre-exponential factor, $k_B$ is the Boltzmann constant, $T$ is the temperature,
and $E_a$ is the activation energy for \ce{Li+} diffusion.
In other words, the logarithm of $D$ exhibits a linear relationship with $1/T$, where the slope is proportional to $E_a$.

Extending the ACE simulations to higher temperatures subsequently revealed that this linear relationship does not describe the entire temperature range for every state point.
A free-breakpoint analysis supports a two-region fit for unstrained \pmtmo with $T_c=382$~K, whereas the available temperature grid for unstrained \pnma does not reliably distinguish a two-region model from a single Arrhenius line.
Accordingly, the zero-strain curves in \fref{fig:sigma:vs:T} and \fref{fig:IC:strain} use the same diffusivity data and fitting procedure.
We found no experimental measurements that extend through the high-temperature regime and therefore restrict the comparison to the available low-temperature data.
Consequently, the fitted crossover temperature $T_c$ and the associated high-temperature behavior are simulation predictions that await experimental validation.

The room-temperature conductivities collected in \tref{tab:sigma:Ea} span more than three orders of magnitude, but the experimental and the computational values contribute to this spread in different ways.
The experimental values differ among themselves by less than an order of magnitude, largely through the degree of cation site disorder set by the synthesis route: mechanochemical milling of the same composition with and without subsequent annealing yields different conductivities~\cite{asano2018solid}, and the milling conditions can be used deliberately to tune this disorder~\cite{schlem2020mechanochemical}.
They also lie below the computational values, because impedance measurements on polycrystalline pellets include grain-boundary and porosity contributions that are absent from an ideal crystal; the measurement provenance of each reported value is summarized in Sec.~S4 of the \sishort.
The computational values vary far more, dominated by the exchange--correlation functional, which within a single study shifts the room-temperature estimate from 152.66 to 0.56~mS/cm~\cite{qi2021bridging}, and by extrapolation of high-temperature AIMD to room temperature with a single Arrhenius line (values marked $\dagger$)~\cite{hu2023revealing,park2020theoretical,wang2019lithium,chen2024accelerating}.

\begin{figure}[tbh!]
  \centering
  \includegraphics[width=\linewidth]{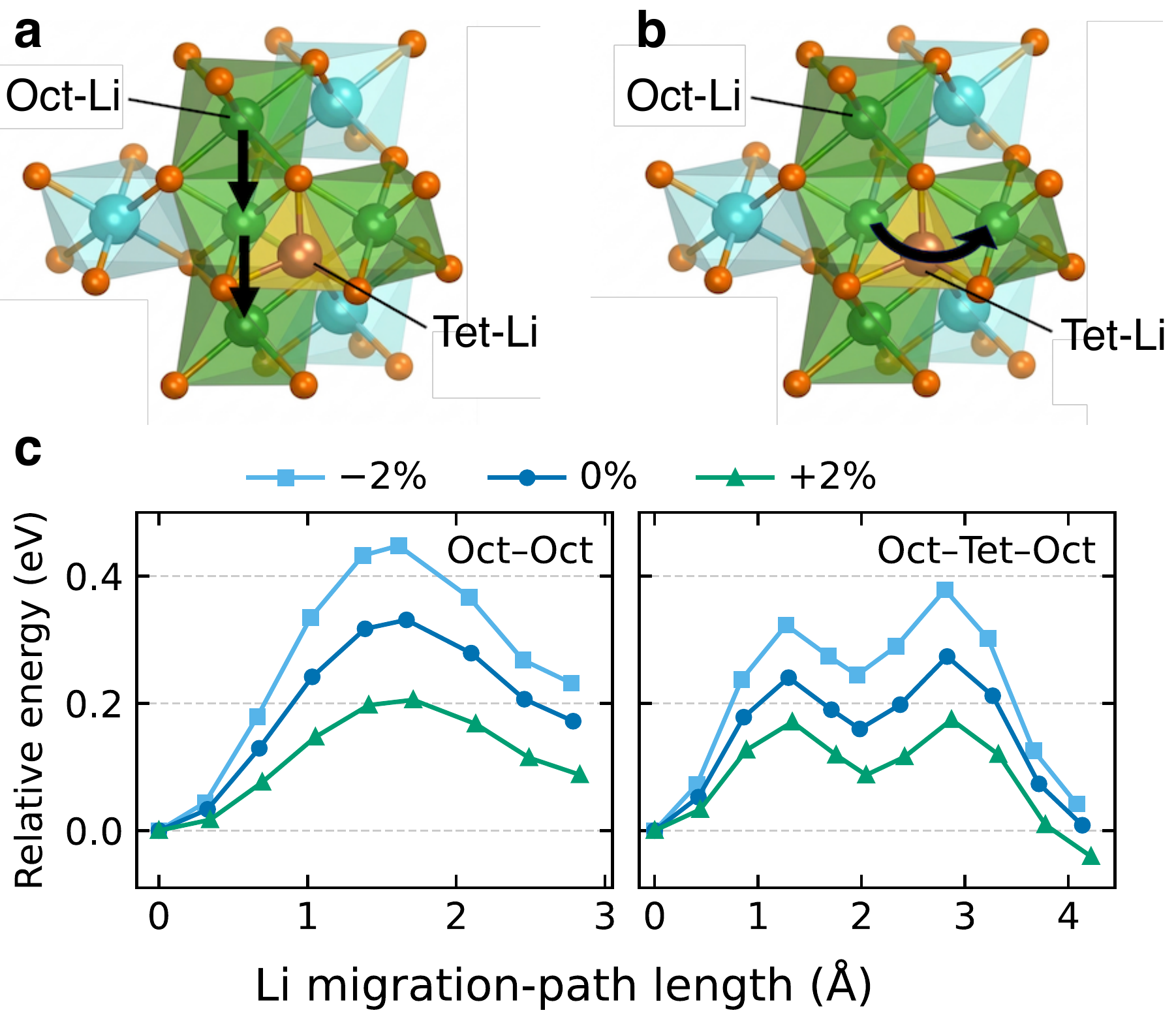}
  \caption{\textbf{\ce{Li+} diffusion pathways and their migration-energy profiles in the \pmtmo phase.}
    Diffusion pathways along the (a) Oct--Oct sites and (b) Oct--Tet--Oct sites.
    (c) Migration-energy profiles of the two pathways, obtained with the final ACE model using the climbing-image nudged elastic band method for a single migrating \ce{Li+} at fixed-cell isotropic strains of $-2$\%, 0\%, and $+2$\%.
    Energies are relative to the initial Oct site of each path.
    The local minimum near 2~\AA\ in the Oct--Tet--Oct profile is the Tet intermediate site.
  }
  \label{fig:diff:pathway}
\end{figure}

Where statistically resolved, the simulated change in Arrhenius slope is accompanied by a change in the accessibility and dimensionality of the \ce{Li+} diffusion pathways.
This interpretation is consistent with prior simulations~\cite{hu2023revealing,wang2023frustration,schlem2021insights}.
More broadly, continuous superionic crossovers with a change in Arrhenius slope are historically described as Faraday transitions in type-II superionic conductors.
Besides $\beta$-\ce{PbF2}, examples include fluorite-structured \ce{CaF2}, \ce{SrF2}, \ce{BaF2}, and \ce{SrCl2}, as well as antifluorite \ce{Li2O}~\cite{ishii2001cluster,fossati2021structural}.
In these materials, conductivity increases continuously over a finite temperature interval without an abrupt transformation of the crystalline framework; in archetypal fluorites, this crossover is associated with increasing disorder of the mobile-anion sublattice and a change from isolated-defect hopping to correlated motion~\cite{annamareddy2017low,fossati2021structural}.
The crossover in \lyc shares these continuous transport signatures.
Microscopically, however, it reflects the increasing accessibility of intrinsic \ce{Li+} sites and pathways rather than the anion Frenkel disorder found in fluorite compounds.

\ce{Li+} migration in \ce{Li3YCl6} is governed by two intrinsic diffusion pathways: (1) the direct octahedral--octahedral (Oct--Oct) hop that requires no intermediate sites (\fref{fig:diff:pathway}a), and (2) the octahedral--tetrahedral--octahedral (Oct--Tet--Oct) hop that proceeds through the tetrahedral intermediate site (\fref{fig:diff:pathway}b).
To quantify the two pathways, we computed static migration-energy profiles at zero temperature for a single migrating \ce{Li+} in the fixed anion framework, using the climbing-image nudged elastic band (NEB) method and following the approach of \olcite{wang2023design} (\fref{fig:diff:pathway}c).
The Oct--Tet--Oct path proceeds through a Tet intermediate that lies 0.160~eV above the Oct site.
Tension lowers both the barriers and the Tet site energy, and compression raises them.

\begin{figure}[tbh!]
  \centering
  \includegraphics[width=0.8\linewidth]{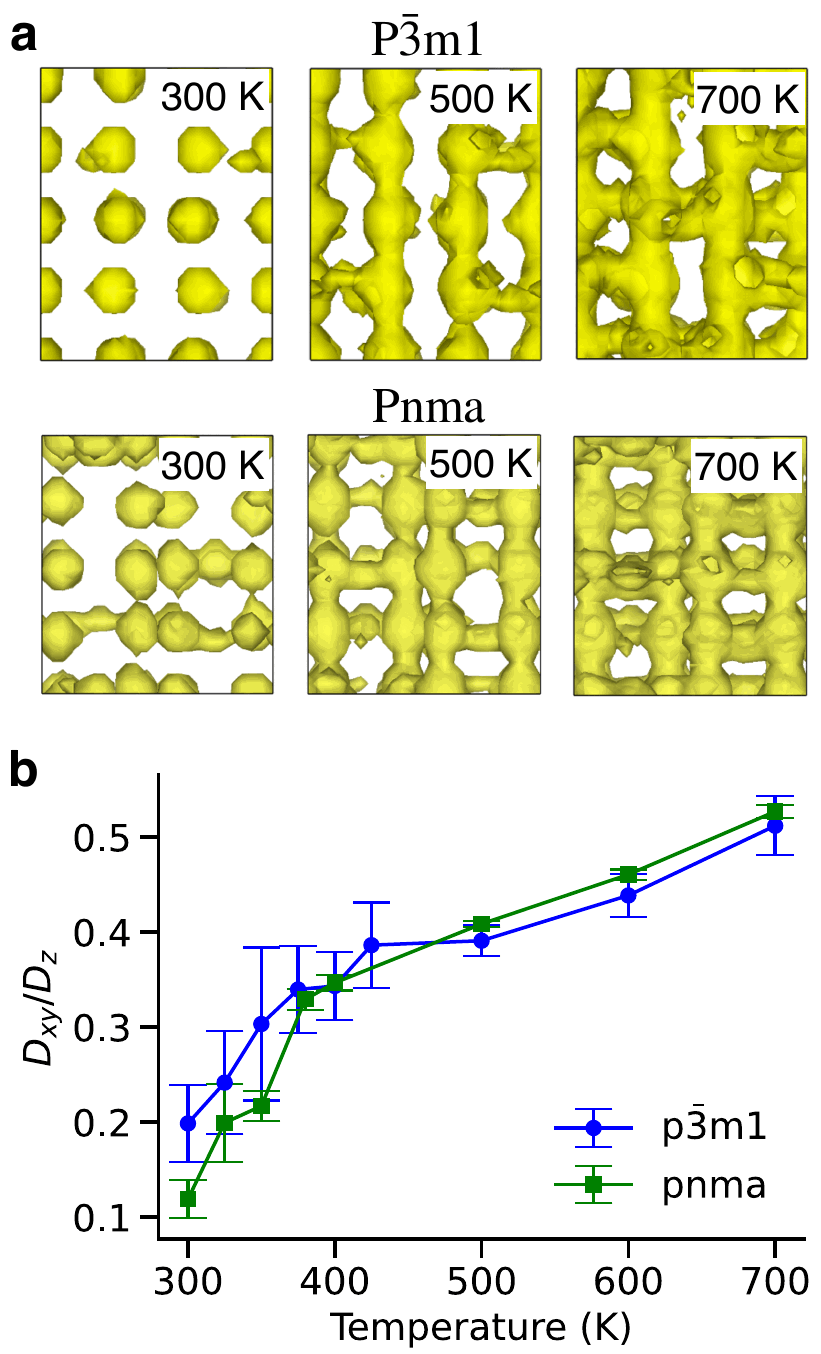}
  \caption{\textbf{Contribution of different diffusion pathways to the overall \ce{Li+} transport}.
    (a) \ce{Li+} probability density showing the trajectory of \ce{Li+} ions during MD simulations, viewed from the $b$-$c$ plane in \fref{fig:structure}.
    (b) Ratio of the diffusivity along the Oct--Tet--Oct pathway $D_{xy}$ to that along the Oct--Oct pathway $D_z$.
  }
  \label{fig:occupancy:zero:strain}
\end{figure}

The behavior at finite temperature is resolved directly by the \ce{Li+} probability density map, which shows the likelihood of finding an \ce{Li+} ion at each location in the simulation cell along the MD trajectory (generation method given in Sec.~S4 of the \sishort).
At 300~K, the map for \pmtmo in \fref{fig:occupancy:zero:strain}a shows \ce{Li+} density concentrated almost entirely at the Oct sites.
Vertical Oct--Oct hopping is already active at this temperature, as evidenced by the nonzero diffusivity in \fref{fig:sigma:vs:T}, but the events are too infrequent to produce visible density in the map.
As the temperature increases to 500~K, continuous one-dimensional channels emerge along the vertical direction via Oct--Oct hopping, and horizontal Oct--Tet--Oct bridges also appear.
Although less pronounced than the vertical channels, they are already sufficient to link the 1D channels into a three-dimensional network.
At 700~K, the horizontal Oct--Tet--Oct pathways become more pronounced and contribute an increasingly large share of the overall \ce{Li+} transport.

To quantitatively assess the relative contributions of the two pathways, we compute $D_{xy}$ and $D_z$ by isolating the $xy$ and $z$ components of the \ce{Li+} diffusivity, respectively, and plot their ratio in \fref{fig:occupancy:zero:strain}b.
The ratio rises monotonically from ${\sim}0.2$ at 300~K to ${\sim}0.55$ at 700~K, confirming that the thermally activated opening of Oct--Tet--Oct pathways progressively enhances the transverse contribution to \ce{Li+} diffusion.

This ratio is at first sight difficult to reconcile with the static profiles of \fref{fig:diff:pathway}c, which place the Oct--Tet--Oct barrier below the Oct--Oct barrier and would therefore predict a transverse contribution that dominates at every temperature.
Those profiles, however, describe a single ion migrating through a rigid lattice at zero temperature, and omit both the thermal population of the Tet sites and the local rearrangement required to admit a \ce{Li+} into one.
The quantity that governs transport at finite temperature is instead the free-energy barrier, which we extracted path-resolved directly from the MD trajectories by Boltzmann inversion of the \ce{Li+} probability density, as described in Sec.~S4 of the \sishort.
On this measure the Oct--Tet--Oct barrier exceeds the Oct--Oct barrier in \pmtmo at every temperature and strain examined, so transport at low temperature is predominantly associated with Oct--Oct hopping~\cite{lee2024disorder}.
As the Tet sites populate, the two barriers become comparable in thermal units, consistent with the rise of $D_{xy}/D_z$.

We hypothesize that the temperature-dependent accessibility of Tet sites originates in the flexibility of the \ce{Cl-} anion lattice: a stiff lattice at low temperature restricts the local distortion required to form a Tet intermediate site, whereas the larger thermal vibrations at higher temperature open transient gaps between the \ce{Cl-} ions.
To verify this hypothesis, we measure the \ce{Cl-} response directly.
The \ce{Cl-} ions vibrate about fixed sites, and the amplitude of that vibration grows from about 6\% of the distance to their nearest \ce{Cl-} neighbors at 300~K to 11\% at 800~K in both phases, so the anion framework does loosen over the range in which the crossover occurs (Sec.~S7 of the \sishort).
Strain acts in the same direction at fixed temperature: dilating the framework raises the Tet fraction, whereas compressing it lowers the fraction (Table~S14 of the \sishort).
These observations support the flexibility hypothesis: the anion framework loosens and the Tet population grows together under both temperature and strain.

When the simulations are extended to high temperature, the model trained on optB88-vdW data predicts two Arrhenius regions for several phase and strain states, whereas those trained on PBEsol and on PBEsol+MBD data remain approximately single-regime over the sampled range (Fig.~S6 in the \sishort).
Adding the many-body dispersion treatment to PBEsol therefore does not by itself produce the crossover.
Because the three treatments differ in exchange and local correlation as well as in dispersion, the comparison demonstrates sensitivity to the complete exchange--correlation treatment rather than identifying vdW interactions as the microscopic origin of the crossover.

This functional sensitivity helps explain the broad variation among simulated ionic conductivities $\sigma$ reported for \ce{Li3YCl6}.
AIMD simulations are typically performed at high temperatures to accelerate \ce{Li+} diffusion and obtain sufficient statistics within computationally feasible timescales.
The room-temperature $\sigma$ is then extrapolated from the high-temperature data using a single linear Arrhenius relationship~\cite{hu2023revealing,park2020theoretical,wang2019lithium,chen2024accelerating}, which can miss a simulated change in slope and thereby produce a different low-temperature extrapolation (see \tref{tab:sigma:Ea}).
Classical MD simulations using interatomic potentials fitted to PBE or PBEsol data remain approximately single-regime over the sampled temperature range, reflecting the sensitivity of the predicted temperature dependence to the underlying exchange--correlation treatment.
They overestimate $\sigma$ at room temperature as well, as seen in the literature MLIP fitted to PBE data~\cite{qi2021bridging} (see \tref{tab:sigma:Ea}) and in our ACE models trained on PBEsol and PBEsol+MBD data (Table~S7 in the \sishort).

\subsection{Strain Effect}
\label{sec:strain}

To investigate the effects of strain on the \ce{Li+} transport behavior in \ce{Li3YCl6}, we perform MD simulations under isotropic compressive and tensile strains ranging from $-2$\% to 2\%.
For a target strain $\varepsilon$, all three cell vectors were scaled by $1+\varepsilon$ while the fractional atomic coordinates were retained.
The strained cell was then held fixed and the internal atomic coordinates were left free, both in the NVT MD simulations and in the structure optimizations carried out under strain.
Assuming linear elasticity and using the elastic tensors computed in \sref{sec:validation}, a 2\% isotropic strain corresponds to a maximum stress of about 1.7~GPa for both the \pmtmo and \pnma phases (calculation given in Sec.~S6 of the \sishort).
This nominal stress exceeds the typical stack pressure applied in experiments, which is on the order of tens to hundreds of MPa~\cite{krauskopf2019toward,yang2021interfacial,zhao2023taming}.
The actual internal stress can be significantly higher, however, because of interfacial roughness, microstructural heterogeneity, and stress concentrations~\cite{doux2020pressure,janek2023challenges}, which makes the selected strain range relevant to practical battery conditions.

\begin{figure}[tbh!]
  \centering
  \includegraphics[width=0.9\columnwidth]{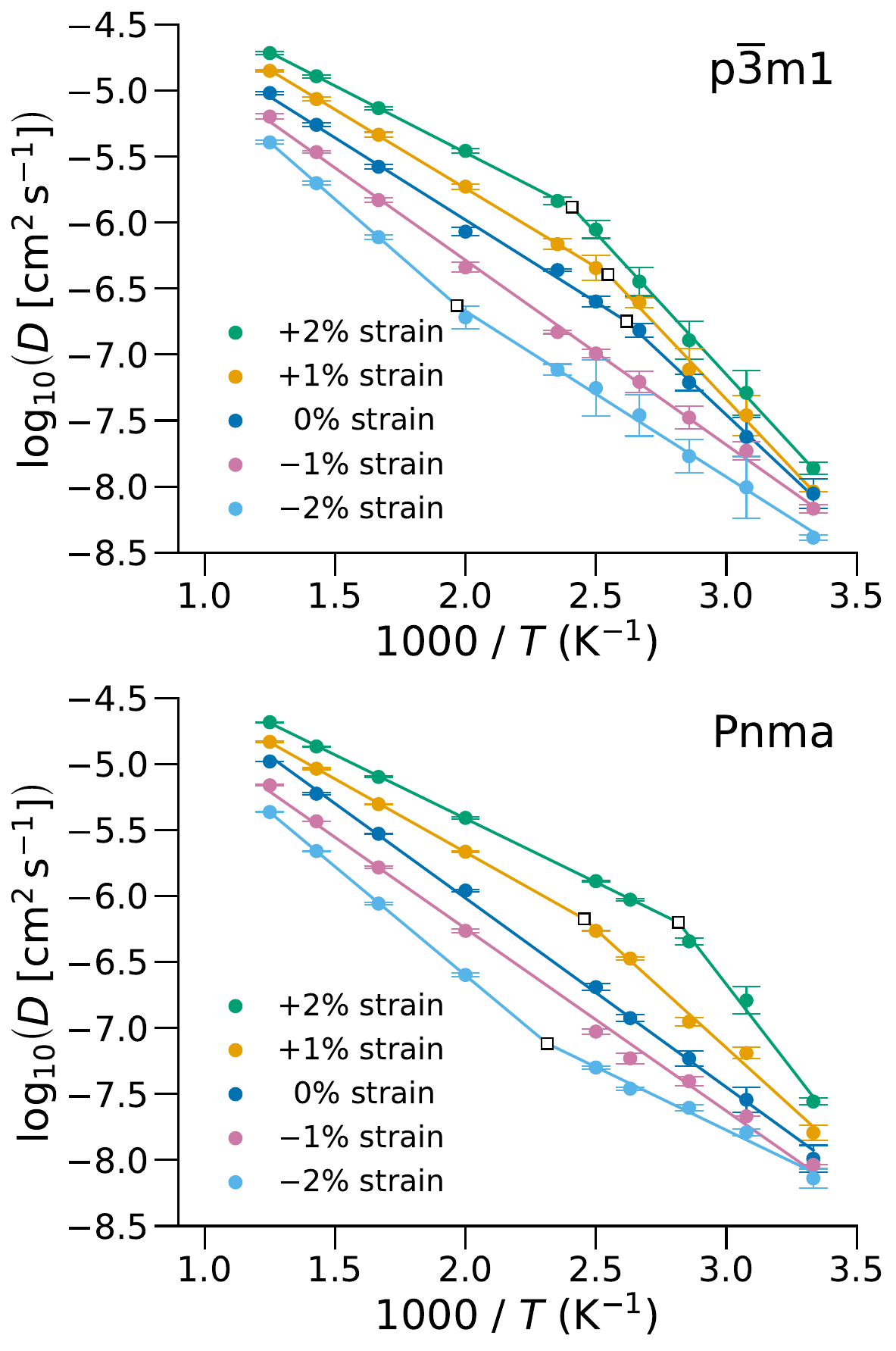}
  \caption{\textbf{Arrhenius plot of the Li diffusivity $D$ of \ce{Li3YCl6} under strain.}
    Shown for strain conditions ranging from $-2$\% to $+2$\%, computed with the final ACE model.
    Open black squares mark the fitted crossover temperatures $T_c$ for the continuous two-region fits.
    Error bars represent standard deviations from three independent MD runs.
  }
  \label{fig:IC:strain}
\end{figure}

The effect of strain on the diffusivity is shown in \fref{fig:IC:strain}.
Positive tensile strains shift the curves upward compared to the unstrained case across the entire temperature range, indicating higher diffusivity $D$ and ionic conductivity $\sigma$.
Conversely, negative compressive strains shift the curves downward, leading to lower $D$ and $\sigma$ (values given in \tref{tab:strain:sigma:Ea}).
Free-breakpoint fits and AICc model selection resolve two Arrhenius regions for seven of the ten phase/strain states, with $T_c$ ranging from 355 to 508~K (\tref{tab:strain:sigma:Ea}).
A single Arrhenius line is used for \pmtmo at $-1$\% strain and \pnma at $-1$\% and 0\% strain because the sampled temperature grid does not reliably resolve a crossover for these states.
Removing the sampled temperature nearest the formerly assumed boundary changes the retained $T_c$ values by at most 9~K, although the model preference weakens in some cases.
Thus, the data do not support a strain-invariant $T_c$, but they continue to show the same Oct--Oct and Oct--Tet--Oct transport framework across the strain range.
However, the slope of the Arrhenius curves in the two regimes is modified in a different manner, indicating that strain modulates the efficiency and accessibility of the diffusion pathways in a temperature-dependent manner.

\begin{table}[tbh!]
  \centering
  \caption{\textbf{Strain-dependent ionic conductivity and activation energy.}
    Ionic conductivity $\sigma$ is reported in mS/cm at $T=300$~K.
    The fitted $T_c$ and activation energies $E_a$ (in eV) are reported for the model selected by AICc.
    A dash denotes a single-line Arrhenius fit, for which the single $E_a$ spans both activation-energy columns.
  }
  \label{tab:strain:sigma:Ea}
  \begin{tabular}{cccccc}
    \hline
                            & Strain & $\sigma$ & $T_c$ (K) & $E_a^\mathrm{low}$ & $E_a^\mathrm{high}$ \\
    \hline
    \multirow{5}{*}{\pmtmo} & $-2$\% & 0.38 & 508 & 0.25 & 0.34 \\
                            & $-1$\% & 0.61 & -   & \multicolumn{2}{c}{0.28 (single)} \\
                            & 0\%    & 0.77 & 382 & 0.37 & 0.25 \\
                            & $+1$\% & 0.78 & 393 & 0.41 & 0.24 \\
                            & $+2$\% & 1.13 & 415 & 0.43 & 0.20 \\
    \hline
    \multirow{5}{*}{\pnma}  & $-2$\% & 0.66 & 432 & 0.19 & 0.33 \\
                            & $-1$\% & 0.82 & -   & \multicolumn{2}{c}{0.27 (single)} \\
                            & 0\%    & 0.88 & -   & \multicolumn{2}{c}{0.28 (single)} \\
                            & $+1$\% & 1.36 & 407 & 0.36 & 0.22 \\
                            & $+2$\% & 2.25 & 355 & 0.51 & 0.19 \\
    \hline
  \end{tabular}
\end{table}

Consistent with Fig.~S7 in the \sishort, the static energy profile of the representative Oct--Oct hop becomes flatter under tensile strain.
These single-event potential-energy barriers should not be equated with the Arrhenius activation energies in \tref{tab:strain:sigma:Ea}, which are effective finite-temperature quantities that reflect the ensemble of accessible pathways and their populations.

\begin{figure*}[tbh!]
  \centering
  \includegraphics[width=\linewidth]{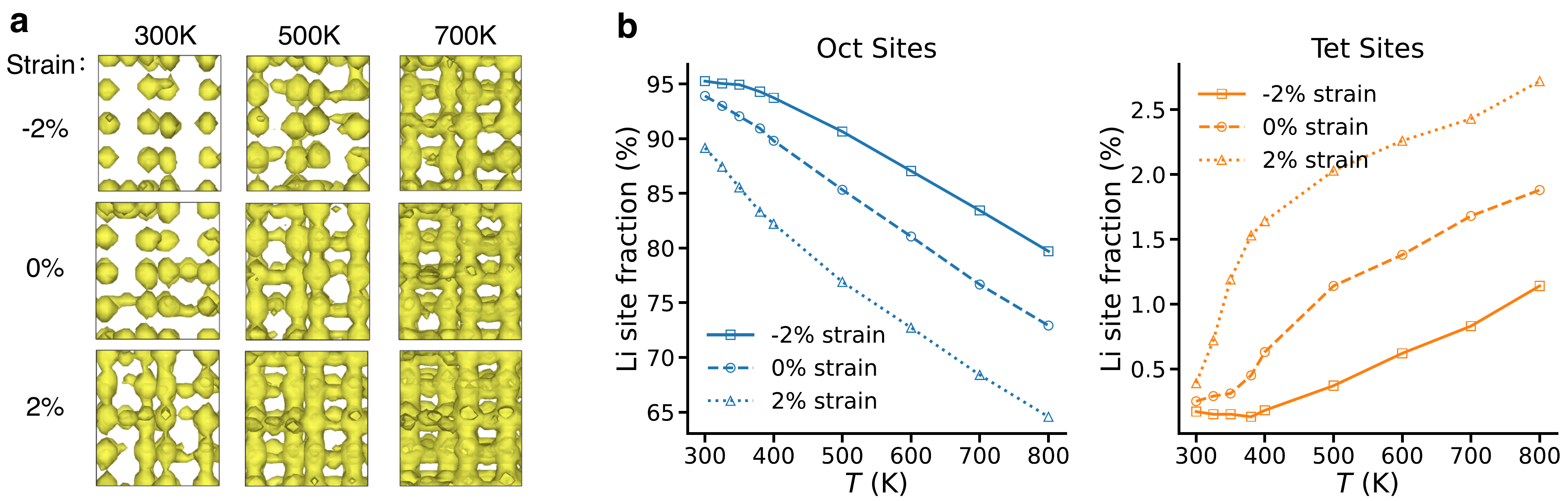}
  \caption{\textbf{Effect of strain on the \ce{Li+} transport behavior in \pnma \lyc.}
    (a) \ce{Li+} probability density at different strains and temperatures, viewed from the $b$-$c$ plane in \fref{fig:structure}.
    (b) \ce{Li+} occupancy fraction for the octahedral (Oct) and tetrahedral (Tet) sites at different temperatures.
  }
  \label{fig:strain:effects}
\end{figure*}

In the low-temperature regime $T < T_c$, tensile strain induces a counterintuitive trend: the activation energy $E_a$ increases with the strain magnitude, while the diffusivity $D$ remains higher than the unstrained case (\fref{fig:IC:strain}).
The fitted prefactors are reported in Table~S11 in the \sishort.
For a two-region fit, $D_0$ is not independent of $E_a$: the two branches meet at the breakpoint, so the high-temperature intercept is fixed by the corresponding slope and by the breakpoint diffusivity.
We therefore report $D_0$ as a fit parameter and do not assign its variation across strain states to independent microscopic contributions.
As discussed in \sref{sec:li:transport}, the dominant diffusion pathway in the low-temperature regime for unstrained \lyc is the direct Oct--Oct hop; the multistep Oct--Tet--Oct pathway is restricted due to the high formation energy of Tet intermediate sites.
Tensile strain leads to lattice expansion and thus reduces the formation energy of Li occupying the Tet sites, resulting in a significant increase in Tet site occupancy even at low temperatures.
This is directly confirmed by \fref{fig:strain:effects}a: the \ce{Li+} probability density at 300~K becomes progressively more delocalized with increasing tensile strain, with additional horizontal Oct--Tet--Oct diffusion channels emerging at 2\% strain.
\fref{fig:strain:effects}b quantifies this further using the \ce{Li+} site occupancy fraction (calculation details in Sec.~S4 of the \sishort): at 300~K, 2\% tensile strain produces a $\sim3\times$ increase in Tet site occupancy relative to the unstrained case, while Oct site occupancy decreases by $\sim$6\%.
The tensile-strain enhancement is robust to the coordination cutoff (Sec.~S4 of the \sishort).
The activated Oct--Tet--Oct pathways have a higher intrinsic activation energy than the direct Oct--Oct pathway, thus raising the overall $E_a$.
The accompanying increase in $D$ therefore cannot be inferred from $E_a$ alone; the independent occupancy and probability-density analyses instead show that tensile strain makes a broader set of Li configurations and diffusion pathways accessible.

In the high-temperature regime ($T > T_c$), tensile strain enhances \ce{Li+} diffusivity while reducing the activation energy $E_a$.
As discussed in \sref{sec:li:transport}, the high thermal energy at $T > T_c$ makes Tet intermediate sites and the Oct--Tet--Oct pathway more accessible in unstrained \lyc, producing a connected three-dimensional network and a lower effective high-temperature activation energy.
Lattice expansion induced by tensile strain further lowers this intrinsic barrier.
For example, $E_a$ falls from 0.25~eV at 0\% strain to 0.20~eV at 2\% strain for \pmtmo (\tref{tab:strain:sigma:Ea}).
Our atomistic simulation data validate this: \fref{fig:strain:effects}b shows elevated Tet site occupancy under 2\% strain relative to the unstrained system at 700~K, and \fref{fig:strain:effects}a reveals a more open, well-connected 3D \ce{Li+} diffusion network; the same behavior is found for \pmtmo (Table~S14 in the \sishort).
The reduced $E_a$ and increased pathway connectivity are therefore consistent with the strong improvement of $D$ under tensile strain at elevated temperatures.

Under compressive strain, the characteristic upward-bent curves seen in the unstrained and tensile cases are replaced by distinct downward-bent curves (\fref{fig:IC:strain}).
On the fitted low-temperature branch, compressive strain reduces the apparent activation energy $E_a$ (e.g., from 0.37~eV at 0\% strain to 0.25~eV at $-2$\% strain for \pmtmo, \tref{tab:strain:sigma:Ea}), yet significantly suppresses total $D$ relative to the unstrained system.
This counterintuitive combination again shows that the fitted activation energy alone does not determine the magnitude of $D$.
Lattice contraction reduces the accessibility of intermediate Tet sites: their occupancy remains negligible from 300~K to 400~K under $-2$\% strain (\fref{fig:strain:effects}b).
Diffusion therefore occurs predominantly through the Oct--Oct pathway on the sampled timescale, consistent with the localized \ce{Li+} probability density at 300~K under $-2$\% strain in \fref{fig:strain:effects}a.
In \pnma the path-resolved free energies place the Oct--Tet--Oct barrier slightly below the Oct--Oct barrier under compression (Sec.~S4 of the \sishort), but the negligible Tet occupancy leaves that pathway carrying little of the current.
In the high-temperature regime ($T > T_c$), compressive strain reduces the Tet site occupancy and disrupts the percolation of the 3D Oct--Tet--Oct diffusion network ($-2$\% data at 700~K in \fref{fig:strain:effects}a and b).
Consequently, the increased activation energy and reduced pathway connectivity both suppress \ce{Li+} transport.

It is useful to place both the magnitude and the sign of the strain response in the context of other solid electrolytes.
Here, 2\% isotropic tensile strain increases the 300~K Nernst--Einstein conductivity by factors of approximately 1.5 and 2.6 for the \pmtmo and \pnma phases, respectively.
AIMD calculations for $\beta$-\ce{Li3PS4} report larger changes of opposite sign for the same type of deformation: strain states that compress the $c$ axis, including $-2$\% isotropic compression, raise the room-temperature conductivity by factors of 2 to 14, whereas those that stretch it, including $+2$\% isotropic tension, lower it~\cite{vzguns2022strain}.
Elsewhere the sign differs again: compressive strain lowers the \ce{Li+} migration barrier in \ce{Al}-doped \ce{Li7La3Zr2O12} but raises it in \ce{Li10GeP2S12}~\cite{moradabadi2020effect}, while anti-perovskite \ce{Li3OCl} and \ce{Li3OBr} become more conductive under tension~\cite{lin2024strain,lin2024insight}.
No universal rule therefore relates the sign of the strain response to the sense of the deformation; it is set by the host framework and by the pathways that carry the current, which is what the pathway-resolved analysis above establishes for \lyc.

\section{Conclusions}
\label{sec:conclusions}

This work develops an accurate and efficient machine learning interatomic potential (ACE) for the \ce{Li3YCl6} solid-state electrolyte, trained on DFT data using the self-consistent vdW-inclusive optB88-vdW functional.
DFT benchmarks support the model's structural, mechanical, and representative strained migration predictions.
Extending the simulations beyond the temperature range covered by experiment reveals two Arrhenius regions for several phase and strain states, a crossover that depends on the exchange--correlation treatment of the reference data and awaits experimental validation.

Our study reveals two fundamental, mechanistically consistent insights into strain-tunable \ce{Li+} transport in \lyc.
First, free-breakpoint model selection shows that $T_c$ is not strain-invariant: it varies nonmonotonically where two regimes are resolved, and the available temperature grid does not reliably resolve a crossover for three strain states.
Nevertheless, the probability densities, site occupancies, and path-resolved free energies consistently identify the same Oct--Oct and Oct--Tet--Oct transport framework across the strain range.
Second, the mechanism of strain tuning is inherently coupled to the dominant diffusion pathway, and thus strongly temperature-dependent.
Tensile strain increases the accessibility of Li configurations and diffusion pathways and reduces the effective activation energy at high temperatures, whereas compressive strain reduces pathway connectivity and increases the high-temperature activation energy.
At low temperatures, the opposite changes in diffusivity and fitted activation energy show that the magnitude of $D$ cannot be interpreted from $E_a$ alone.

These findings have critical implications for the design and operation of solid-state batteries.
Stresses and strains arise not only from stack pressure but also from electrode volume changes and interfacial reactions during electrochemical cycling.
The simulated strain range corresponds to local strain hotspots rather than to the homogeneous response to nominal stack pressure.
Local compressive strain associated with heterogeneous contact, lattice mismatch, interfacial reactions, or defects can therefore suppress ionic conductivity, whereas local tensile strain can enhance it.
Conversely, our results suggest that controlled tensile strain in the electrolyte, while challenging to implement in practice, could enhance \ce{Li+} transport.
Taken together, these results provide a pathway-resolved account of how isotropic strain changes ionic transport in \lyc.

\section*{Data and Code Availability}
The dataset used in this work, along with the scripts for model training, molecular dynamics simulations, and data analysis, is publicly available at \url{https://github.com/weifanfrank/mlip_halide_electrolytes}.
The ACE models are trained using \verb|pacemaker|, which can be accessed at \url{https://github.com/ICAMS/python-ace}.

\section*{Acknowledgements}
This work is supported by the National Science Foundation under Grant No.\ 2316667 and the Research Computing Data Core at the University of Houston.
It uses computational resources provided by the Hefei Advanced Computing Center.

%

\end{document}


\title{\Large Supplemental Material:\\
  Strain-Dependent Ionic Transport in \lyc Solid Electrolytes}

\author{Wei-Fan Huang}
\author{Jin Dai}
\affiliation{William A. Brookshire Department of Chemical and Biomolecular Engineering, University of Houston, Houston, TX, 77204, USA}
\author{Jiahui Pan}
\affiliation{Institute of Fundamental and Frontier Sciences, University of Electronic Science and Technology of China, Chengdu, 611731, China}
\author{Mingjian Wen}
\email{mjwen@uestc.edu.cn}
\affiliation{Institute of Fundamental and Frontier Sciences, University of Electronic Science and Technology of China, Chengdu, 611731, China}
\affiliation{Frontier Center of Energy Distribution and Integration, Tianfu Jiangxi Laboratory, Chengdu, 641419, China}

\maketitle

\setcounter{tocdepth}{2}
\tableofcontents
\clearpage

\section{ACE Model Development}

\subsection{Construction of the Trigonal Atomistic Models}

The 6g Li sites were fully occupied, whereas the experimentally reported 50\% occupancy of the six 6h sites was represented by placing three Li atoms on those sites.
This gives $\binom{6}{3}=20$ possible geometric orderings before symmetry reduction.
The configurations were grouped into symmetry-equivalent sets using the \texttt{StructureMatcher} class in pymatgen~\cite{Ong2013}, which reduces the 20 orderings to three symmetry-distinct ones with multiplicities 12, 6, and 2.
One representative of each was fully relaxed with DFT, until the residual forces fell below 0.02~eV/\AA, and the lowest-energy configuration was retained for the subsequent calculations.

\subsection{AIMD Training Data Selection}
\label{cluster:method}

The initial DFT dataset contained 162 relaxed and isotropically strained structures, 692 structures with random atomic displacements, and 1261 configurations selected from AIMD, for a total of 2115 structures across the \pmtmo and \pnma phases.
The AIMD configurations were generated for the \pmtmo phase at 500~K using 60-atom cells and trajectories of 8~ps duration.
The sampled configurations were subsequently evaluated as static single-point calculations, following the procedure described in Section~II\,C of the main text.

\begin{figure}[tbh!]
  \centering
  \includegraphics[width=0.95\textwidth]{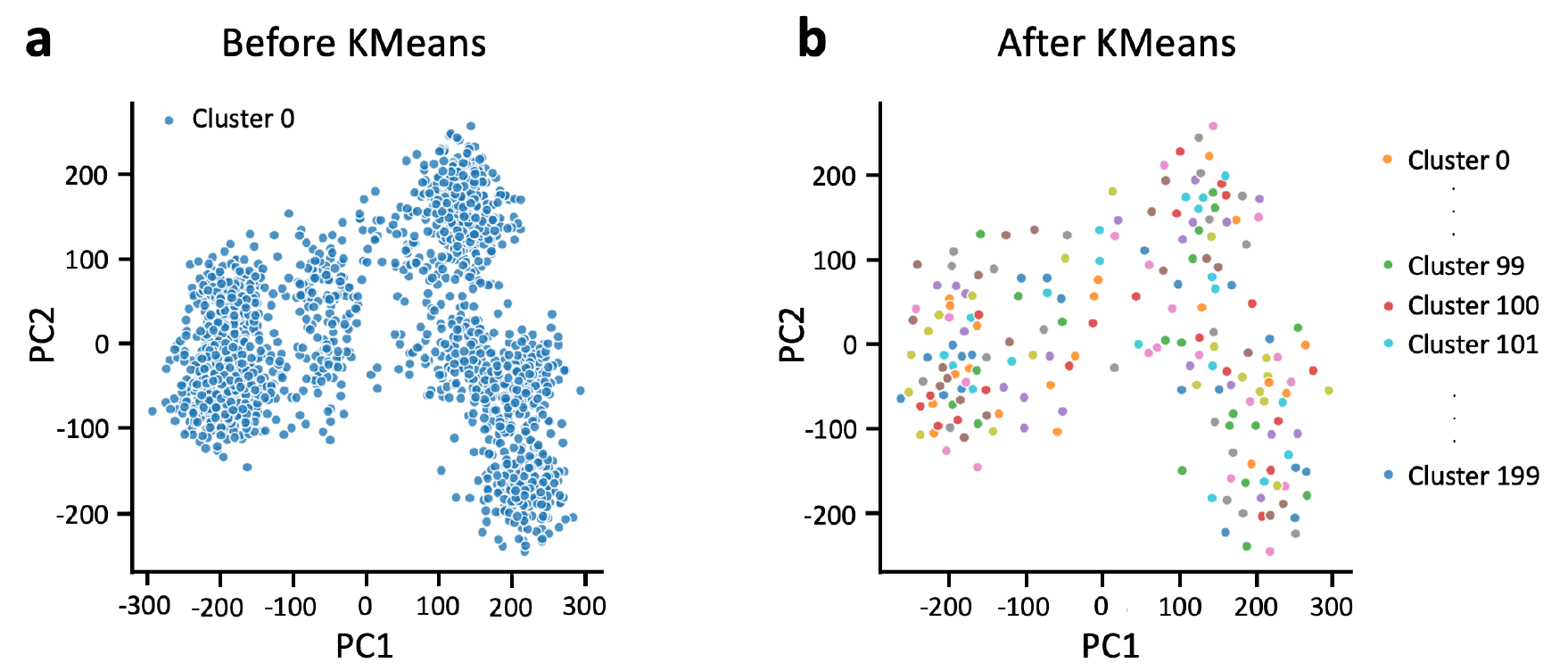}
  \caption{\textbf{Data selection using KMeans clustering.}
    One \pmtmo AIMD trajectory pool at 500~K, shown (a) before and (b) after KMeans clustering.}
  \label{fig:kmeans}
\end{figure}

Since successive snapshots in AIMD trajectories are highly correlated,
we subsampled the trajectory to retain only representative configurations.
To this end, we used DScribe~\cite{himanen2020dscribe} to convert each snapshot
into a Smooth Overlap of Atomic Positions (SOAP)~\cite{bartok2013representing}
vector, which encodes the local atomic environment in a continuous and
invariant representation.
KMeans clustering was applied separately to multiple AIMD trajectory pools in SOAP space.
\fref{fig:kmeans} illustrates one \pmtmo pool, from which 200 representative configurations were retained, with PCA used to reduce the dimensionality for visualization~\cite{roweis1997algorithms}.
Across all trajectory pools, 1261 AIMD configurations were included in the initial dataset.

\subsection{Extrapolation Grade}

The ACE energy of a configuration is a sum of contributions from its local atomic environments, and the model contains $m$ linear parameters $\boldsymbol{\theta}$.
Linearizing the fitting equations about the optimized parameters $\hat{\boldsymbol{\theta}}$, each local atomic environment $\mathfrak{n}$ contributes one equation, whose coefficients form the row vector
\begin{equation}
  \mathbf{b}(\mathfrak{n}) =
  \left(
    \frac{\partial V(\hat{\boldsymbol{\theta}}, \mathfrak{n})}{\partial \theta_1}
    \;\cdots\;
    \frac{\partial V(\hat{\boldsymbol{\theta}}, \mathfrak{n})}{\partial \theta_m}
  \right),
\end{equation}
where $V$ is the local energy.
Stacking the rows of all environments in the training set gives a tall matrix with $m$ columns, in which every row records how one environment constrains the model parameters.

Following the D-optimality criterion~\cite{podryabinkin2023mlip}, the \emph{active set} is the subset of $m$ training environments whose rows form a square matrix $\mathbf{A}$ with maximal $|\det\mathbf{A}|$, found with the MaxVol algorithm.
Geometrically, these are the $m$ rows spanning the largest volume, so the active set collects the most diverse and most extreme environments in the training set, as seen by the model being fitted.

A new environment $\mathfrak{n}_i$ is then compared with the active set by expressing its row as a linear combination of the active-set rows,
\begin{equation}
  \mathbf{c}(\mathfrak{n}_i) = \mathbf{b}(\mathfrak{n}_i)\mathbf{A}^{-1},
\end{equation}
and the extrapolation grades of the environment and of the whole configuration $x$ are
\begin{equation}
  \gamma(\mathfrak{n}_i) = \max_j |c_j(\mathfrak{n}_i)|,
  \qquad
  \gamma(x) = \max_i \gamma(\mathfrak{n}_i).
\end{equation}
A configuration is thus graded by its single worst environment, so that a locally unfamiliar region is not masked by the remaining atoms.

Physically, $\gamma$ measures how far an environment lies outside what the training data already cover.
If $\gamma(\mathfrak{n}_i)\leq1$, the row lies within the volume spanned by the active set, and the predicted energy is a bounded combination of environments the model has been fit to, that is, an interpolation.
If $\gamma(\mathfrak{n}_i)>1$, the environment lies outside that volume and would increase $|\det\mathbf{A}|$ by a factor $\gamma$ if it replaced the corresponding active-set environment; it is more extreme than anything the model has seen, and the prediction is an extrapolation whose severity $\gamma$ quantifies.
Evaluating $\gamma$ requires only the gradient row $\mathbf{b}$ and the precomputed $\mathbf{A}^{-1}$, with no additional DFT calculation, and is therefore cheap enough to be applied to every snapshot along an MD trajectory.

The implementation and active learning workflow used here follow the MLIP formulation~\cite{novikov2020mlip}.
In this work, structures with $\gamma<2$ were retained without labeling, structures with $2\leq\gamma\leq10$ were selected for DFT labeling, and an exploration trajectory was stopped when $\gamma>10$.
The exploration MD used the NVT ensemble at 300, 500, and 700~K and sampled relaxed cells and cells under $\pm1$\% and $\pm2$\% strain for up to 500000 steps per trajectory.
The seven enrichment cycles added 934, 976, 1531, 1073, 636, 1663, and 1134 structures, respectively.
The final dataset contained 10,062 structures, including 6200 (61.62\%) strained configurations.

\subsection{Model Training and Hyperparameter Tuning}

The initial 2115 structures were randomly divided into training, validation, and test subsets in an 8:1:1 ratio.
During each active learning iteration, the newly selected configurations were labeled with DFT, independently divided using the same 8:1:1 ratio, and appended to the corresponding subsets.
The test set therefore expanded throughout active learning and contained held-out configurations from both the initial dataset and every enrichment cycle.
The energy and force errors reported for the final ACE model were evaluated on this expanded test set.
In addition, the snapshots generated during the active learning MD served as an active test set, on which the extrapolation grade was evaluated to assess how well the training data cover the sampled MD configuration space.

The ACE model was trained by minimizing a loss function that balances energy
and force predictions with regularizations that enforce the smoothness of
the potential energy surface~\cite{bochkarev2022efficient}:
\begin{equation}
  \begin{aligned}
    \mathscr{L}    & = (1 - \kappa) \Delta_{E}^{2} + \kappa \Delta_{F}^{2} + \Delta_{coeff} + \Delta_{rad}                                                                                           \\
    \Delta_{coeff} & = L_1 \sum_{p\mu \mathbf{nlL}} \left| c^{\left( p \right)}_{\mu \mathbf{nlL}} \right| + L_2 \sum_{p\mu \mathbf{nlL}} \left| c^{\left( p \right)}_{\mu \mathbf{nlL}} \right|^{2} \\
    \Delta_{rad}   & = \frac{w_0}{r_{\text{cut}}^2} \int r^2 \sum_{\textit{nl}} |R_{\textit{nl}}(r)| \, dr
    + \frac{w_1}{r_{\text{cut}}^2} \int r^2 \sum_{\textit{nl}} \left| \frac{dR_{\textit{nl}}(r)}{dr} \right| \, dr
    + \frac{w_2}{r_{\text{cut}}^2} \int r^2 \sum_{\textit{nl}} \left| \frac{d^2R_{\textit{nl}}(r)}{dr^2} \right| \, dr
  \end{aligned}
\end{equation}
where $\kappa$ controls the relative weight of energy ($\Delta_{E}^{2}$)
and force ($\Delta_{F}^{2}$) errors; $\Delta_{coeff}$ applies $L_1$/$L_2$
regularization on the expansion coefficients $c^{(p)}_{\mu \mathbf{nlL}}$;
and $\Delta_{rad}$ penalizes the zeroth, first, and second derivatives of the
radial basis functions $R_{nl}(r)$ with weights $w_0$, $w_1$, $w_2$,
promoting smoothness up to second order.
Omitting regularization leads to a poor description of high-energy configurations.
The used values are listed in \tref{tab:ace_hyperparameters}.

\begin{table}[tbh!]
  \centering
  \caption{Hyperparameters for the ACE model training.}
  \label{tab:ace_hyperparameters}
  \begin{tabular}{cccccccccc}
    \hline
    $\kappa$ & $L_1$              & $L_2$              & $w_0$              & $w_1$              & $w_2$              & $r_{\text{cut}}$ & \texttt{maxiter} & \texttt{ladder\_step} \\ \hline
    0.35     & $1 \times 10^{-8}$ & $1 \times 10^{-8}$ & $1 \times 10^{-8}$ & $1 \times 10^{-8}$ & $1 \times 10^{-8}$ & 7.000 \text{\AA} & 200              & 12                    \\
    \hline
  \end{tabular}
\end{table}

\begin{figure}[tbh!]
  \centering
  \includegraphics[width=1\textwidth]{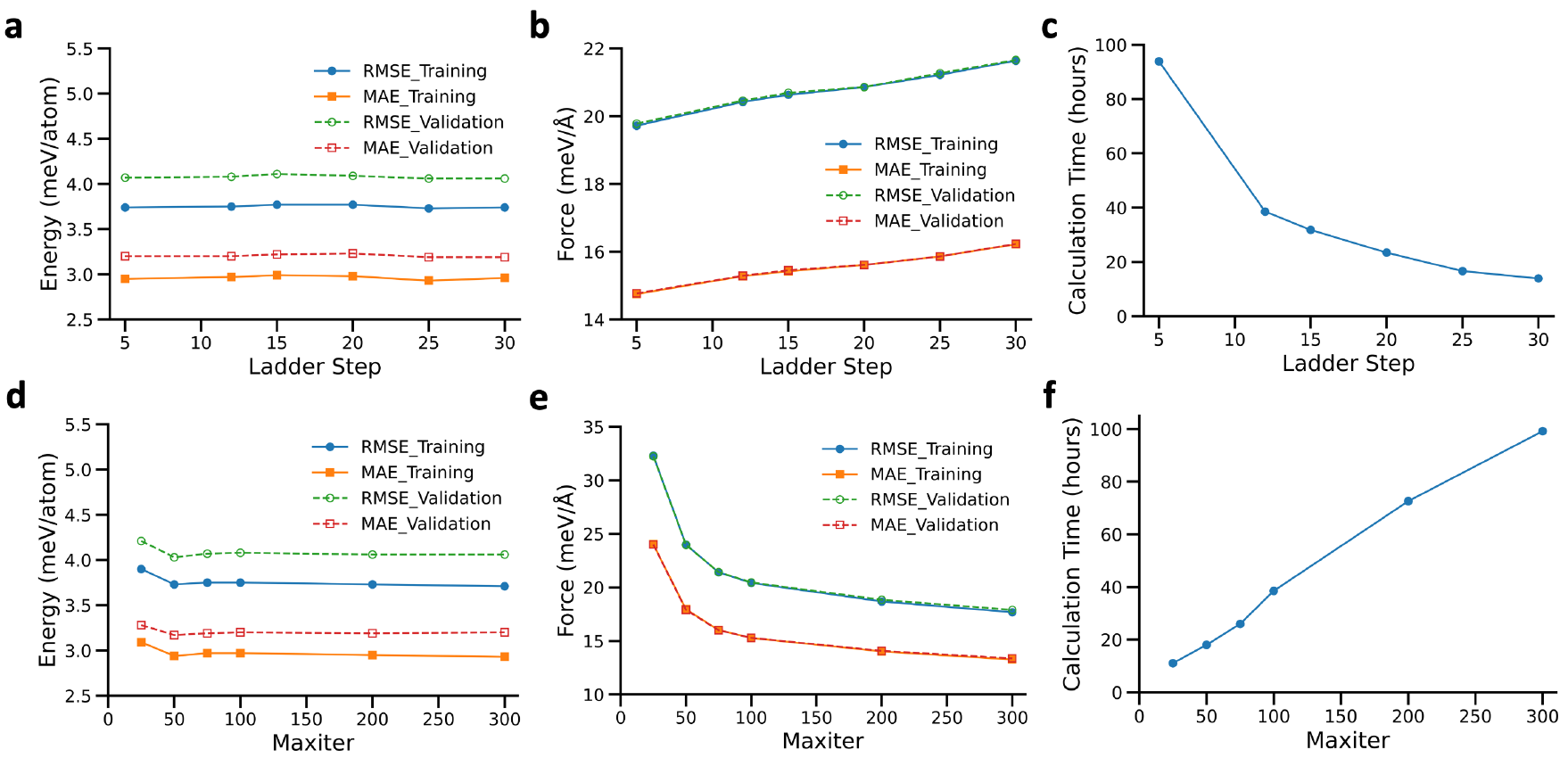}
  \caption{\textbf{Hyperparameter tuning error metrics.}
    (a)--(c) Energy/force RMSE and MAE, and computation time vs.\ \texttt{ladder\_step}
    (\texttt{maxiter} $= 100$).
    (d)--(f) Same metrics vs.\ \texttt{maxiter} (\texttt{ladder\_step} $= 12$).}
  \label{fig:ace:hyperparameter}
\end{figure}

The hyperparameters \verb|maxiter| and \verb|ladder_step| were tuned with
results shown in \fref{fig:ace:hyperparameter}.
\texttt{ladder\_step} controls how many basis functions are added per
iteration: smaller values introduce fewer functions at each step, enabling
more careful optimization at the cost of longer training times, but with
negligible effect on energy errors.
Conversely, increasing \texttt{maxiter} reduces force errors with little
impact on energy accuracy, at the expense of greater computational cost.
A balance between accuracy and efficiency was achieved with
\texttt{maxiter}~$= 200$ and \texttt{ladder\_step}~$= 12$.

\section{Model Performance}

\subsection{optB88-vdW}
\label{ace:potential:optb88vdw}

\begin{table}[tbh!]
  \centering
  \caption{\textbf{Error metrics of 1st to 4th ACE model.}
    Root-mean-square error (RMSE) and mean absolute error (MAE) for energy (meV/atom) and forces (meV/\AA) across training, validation, and test sets.
    Each model corresponds to a different round of active learning (AL).}
  \resizebox{1\textwidth}{!}{
    \label{tab:error:metrics:of:1st:2nd:3rd:and:4th:ace:potential:optb88vdw)}
    \begin{tabular}{cccccccccccccccc}
      \hline
      \multicolumn{1}{c}{{}}      & \multicolumn{3}{c}{{1st (Before AL)}} & \multicolumn{3}{c}{{2nd (After 1st AL)}} & \multicolumn{3}{c}{{3rd (After 2nd AL)}} & \multicolumn{3}{c}{{4th (After 3rd AL)}}                                                                      \\ \hline
      Dataset                     & Metric                                & Energy                                   & Force                                    & Metric                                   & Energy & Force & Metric & Energy & Force & Metric & Energy & Force \\ \hline
      \multirow{2}{*}{Training}   & RMSE                                  & 0.55                                     & 16.02                                    & RMSE                                     & 0.58   & 16.70 & RMSE   & 0.63   & 17.09 & RMSE   & 0.64   & 17.96 \\
                                  & MAE                                   & 0.45                                     & 11.98                                    & MAE                                      & 0.47   & 12.40 & MAE    & 0.51   & 12.57 & MAE    & 0.51   & 13.10 \\ \hline

      \multirow{2}{*}{Validation} & RMSE                                  & 0.59                                     & 16.37                                    & RMSE                                     & 0.61   & 16.75 & RMSE   & 0.64   & 16.98 & RMSE   & 0.65   & 18.01 \\
                                  & MAE                                   & 0.50                                     & 12.20                                    & MAE                                      & 0.50   & 12.40 & MAE    & 0.52   & 12.44 & MAE    & 0.50   & 13.08 \\ \hline
      \multirow{2}{*}{Test}       & RMSE                                  & 0.55                                     & 16.51                                    & RMSE                                     & 0.56   & 17.06 & RMSE   & 0.63   & 17.39 & RMSE   & 0.67   & 18.44 \\
                                  & MAE                                   & 0.46                                     & 12.34                                    & MAE                                      & 0.47   & 12.61 & MAE    & 0.48   & 14.12 & MAE    & 0.53   & 13.48 \\ \hline
    \end{tabular}
  }
\end{table}

\begin{table}[tbh!]
  \centering
  \caption{\textbf{Error metrics of 5th to 8th ACE model.}
    RMSE and MAE for energy (meV/atom) and forces (meV/\AA) across training, validation, and test sets.
    Each model corresponds to a different round of active learning (AL).
    The numbers in parentheses indicate changes relative to the 1st model.}
  \resizebox{1\textwidth}{!}{
    \label{tab:error:metrics:of:5th:6th:7th:and:8th:ace:potential:optb88vdw)}
    \begin{tabular}{ccccccccccccccc}
      \hline
      \multicolumn{1}{c}{{}}      & \multicolumn{3}{c}{{5th (After 4th AL)}} & \multicolumn{3}{c}{{6th (After 5th AL)}} & \multicolumn{3}{c}{{7th (After 6th AL)}} & \multicolumn{3}{c}{{8th (After 7th AL)}}                                                                                    \\ \hline
      Dataset                     & Metric                                   & Energy                                   & Force                                    & Metric                                   & Energy & Force & Metric & Energy & Force & Metric & Energy       & Force         \\ \hline
      \multirow{2}{*}{Training}   & RMSE                                     & 0.62                                     & 18.71                                    & RMSE                                     & 0.63   & 18.73 & RMSE   & 0.70   & 19.24 & RMSE   & 0.67 (+0.08) & 19.18 (+3.16) \\
                                  & MAE                                      & 0.48                                     & 13.69                                    & MAE                                      & 0.48   & 13.70 & MAE    & 0.54   & 14.05 & MAE    & 0.52 (+0.07) & 13.99 (+2.01) \\ \hline
      \multirow{2}{*}{Validation} & RMSE                                     & 0.63                                     & 18.71                                    & RMSE                                     & 0.64   & 18.74 & RMSE   & 0.68   & 19.26 & RMSE   & 0.67 (+0.08) & 19.24 (+2.87) \\
                                  & MAE                                      & 0.46                                     & 13.68                                    & MAE                                      & 0.47   & 13.69 & MAE    & 0.53   & 14.03 & MAE    & 0.51 (+0.01) & 14.02 (+1.82) \\ \hline
      \multirow{2}{*}{Test}       & RMSE                                     & 0.63                                     & 19.03                                    & RMSE                                     & 0.64   & 19.35 & RMSE   & 0.68   & 19.33 & RMSE   & 0.67 (+0.12) & 19.23 (+2.72) \\
                                  & MAE                                      & 0.48                                     & 13.97                                    & MAE                                      & 0.49   & 14.16 & MAE    & 0.53   & 14.13 & MAE    & 0.52 (+0.06) & 14.09 (+1.75) \\ \hline
    \end{tabular}
  }
\end{table}

\begin{figure*}[tbh!]
  \centering
  \includegraphics[width=0.88\textwidth]{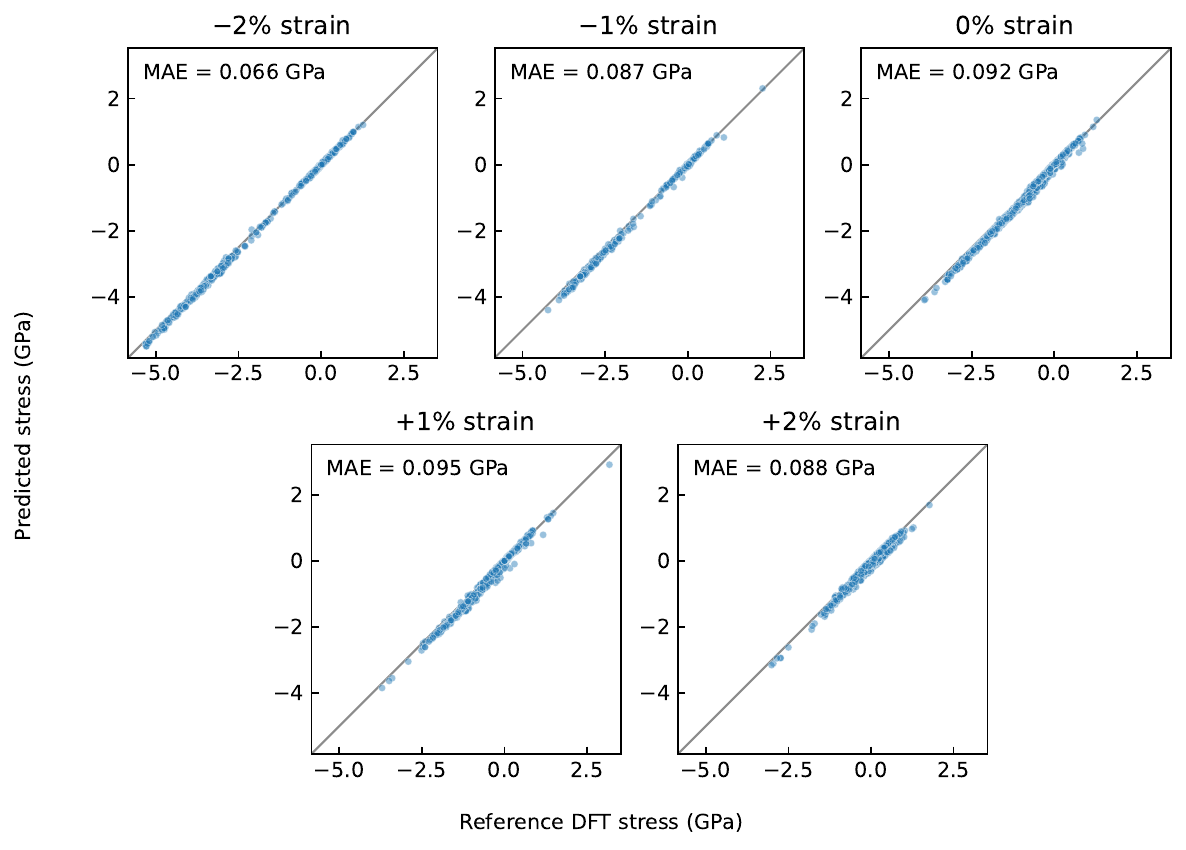}
  \caption{\textbf{DFT validation of ACE stress predictions under strain.}
    Parity plots compare the six stress-tensor components predicted by the final ACE model with the corresponding DFT reference values for held-out structures at $-2$\%, $-1$\%, 0\%, $+1$\%, and $+2$\% isotropic strain.
    All six components are pooled within each panel, and the solid gray line denotes perfect agreement.}
  \label{fig:stress:parity}
\end{figure*}

Although stresses were not included in the fitting loss, the final ACE model was
evaluated against the DFT stresses of the held-out test set.
\fref{fig:stress:parity} compares the six stress-tensor components with their DFT
reference values, and \tref{tab:stress:mae} reports the mean absolute errors
resolved by isotropic strain.
Over the $-2$\% to $+2$\% range used in the transport simulations the overall
error is 0.066--0.095~GPa, without a systematic increase toward the strained
states, and it remains below 0.09~GPa for the $\pm3$\% and $+5$\% structures that
lie outside that range.
The error is carried almost entirely by the normal components; the shear
components agree to 0.012--0.018~GPa at every strain.

\begin{table}[H]
  \centering
  \caption{\textbf{Stress errors of the final ACE model against DFT, resolved by isotropic strain.}
    Mean absolute errors in GPa over the held-out test set, pooled over both phases.
    The normal and shear entries average the three diagonal and the three
    off-diagonal components, respectively.
    Strains beyond $\pm2$\% are not used in the transport simulations and are
    listed to show the behavior outside that range.}
  \label{tab:stress:mae}
  \begin{tabular}{rrccc}
    \hline
    Strain (\%) & Configurations & Overall & Normal & Shear \\
    \hline
    $-3$ & 113 & 0.046 & 0.077 & 0.014 \\
    $-2$ & 117 & 0.066 & 0.117 & 0.014 \\
    $-1$ &  58 & 0.087 & 0.157 & 0.017 \\
    0    & 230 & 0.092 & 0.172 & 0.013 \\
    $+1$ & 105 & 0.095 & 0.174 & 0.016 \\
    $+2$ & 159 & 0.088 & 0.164 & 0.013 \\
    $+3$ & 104 & 0.082 & 0.151 & 0.014 \\
    $+5$ &  44 & 0.048 & 0.085 & 0.012 \\
    \hline
  \end{tabular}
\end{table}

\subsection{PBEsol}
\label{ace:potential:pbesol}

The PBEsol reference calculations followed the same procedure as described in Section~II\,C of the main text, with the exchange--correlation functional replaced by PBEsol~\cite{perdew2008restoring}.

\begin{table}[H]
  \centering
  \caption{\textbf{Error metrics of ACE model trained on DFT data generated with the PBEsol functional.}
    Root-mean-square error (RMSE) and mean absolute error (MAE) for energy (meV/atom) and forces (meV/\AA) across training, validation, and test sets.}
  \label{tab:error:metrics:ace:potential:pbesol}
  \begin{tabular}{cccc}
    \hline
    Dataset                     & Metric & Energy & Force \\
    \hline
    \multirow{2}{*}{Training}   & RMSE   & 0.73   & 20.27 \\
                                & MAE    & 0.59   & 14.78 \\
    \hline
    \multirow{2}{*}{Validation} & RMSE   & 0.71   & 20.18 \\
                                & MAE    & 0.58   & 14.77 \\
    \hline
    \multirow{2}{*}{Test}       & RMSE   & 0.71   & 20.18 \\
                                & MAE    & 0.58   & 14.77 \\
    \hline
  \end{tabular}
\end{table}

\begin{figure}[H]
  \centering
  \includegraphics[width=0.8\textwidth]{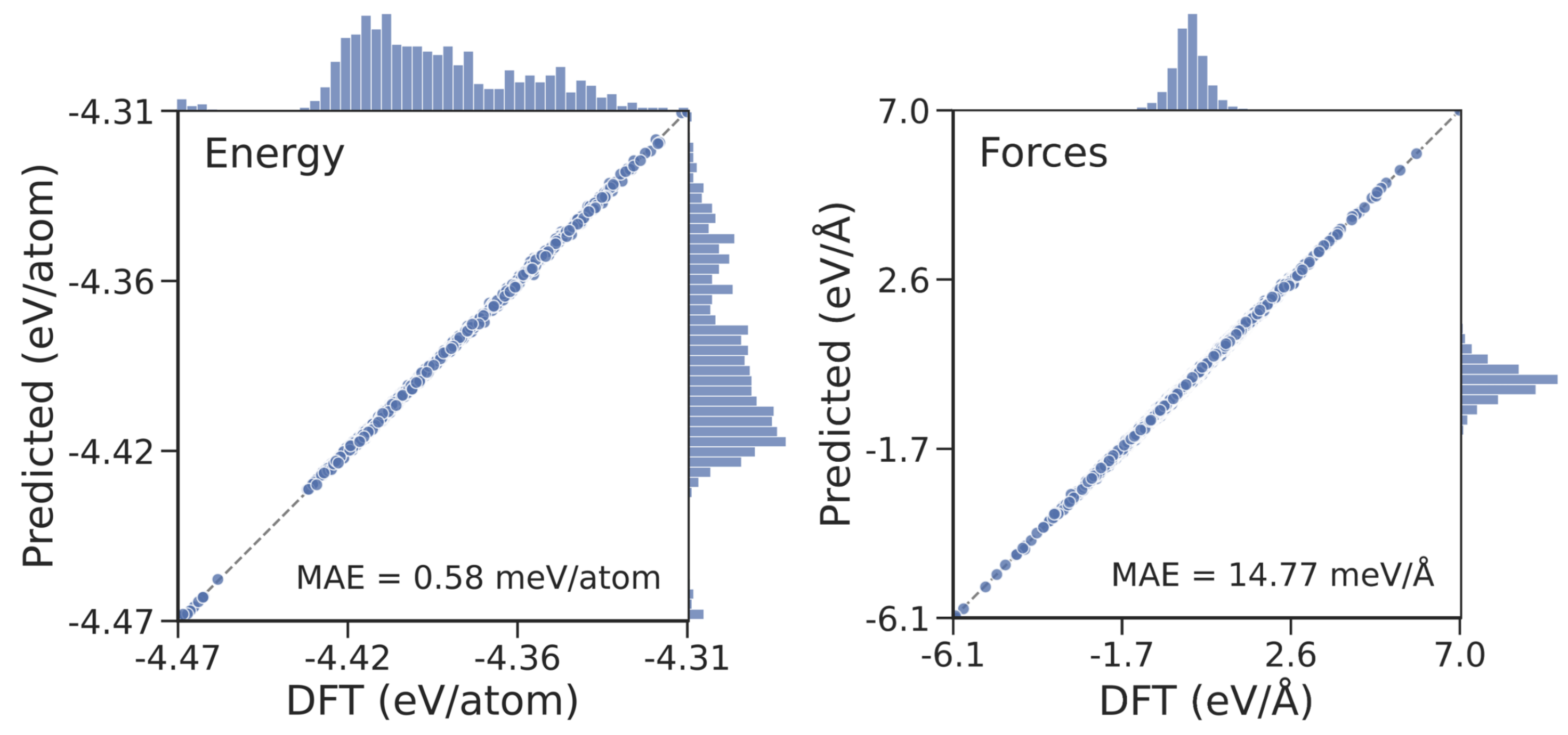}
  \caption{\textbf{Comparison between DFT (PBEsol) and ACE model predictions for energy and forces on the test set.}
    The ACE model was trained on data generated with the PBEsol functional.}
  \label{fig:comparison:of:dft:and:ace:potential:pbesol}
\end{figure}

\subsection{PBEsol+MBD}
\label{ace:potential:MBD}

The PBEsol+MBD reference calculations followed the same procedure, with the exchange--correlation functional replaced by PBEsol combined with the fractionally ionic many-body dispersion method with range-separated self-consistent screening (MBD@rsSCS/FI) of Gould \emph{et al.}~\cite{gould2016fractionally}.

\begin{table}[H]
  \centering
  \caption{\textbf{Error metrics of ACE model trained on DFT data generated with the PBEsol with MBD functional.}
    Root-mean-square error (RMSE) and mean absolute error (MAE) for energy (meV/atom) and forces (meV/\AA) across training, validation, and test sets.}
  \label{tab:error:metrics:ace:potential:mbd}
  \begin{tabular}{cccc}
    \hline
    Dataset                     & Metric & Energy & Force \\
    \hline
    \multirow{2}{*}{Training}   & RMSE   & 3.73   & 20.27 \\
                                & MAE    & 2.95   & 14.02 \\
    \hline
    \multirow{2}{*}{Validation} & RMSE   & 4.06   & 18.84 \\
                                & MAE    & 3.19   & 14.07 \\
    \hline
    \multirow{2}{*}{Test}       & RMSE   & 3.93   & 18.78 \\
                                & MAE    & 3.02   & 14.04 \\
    \hline
  \end{tabular}
\end{table}

\begin{figure}[H]
  \centering
  \includegraphics[width=0.8\textwidth]{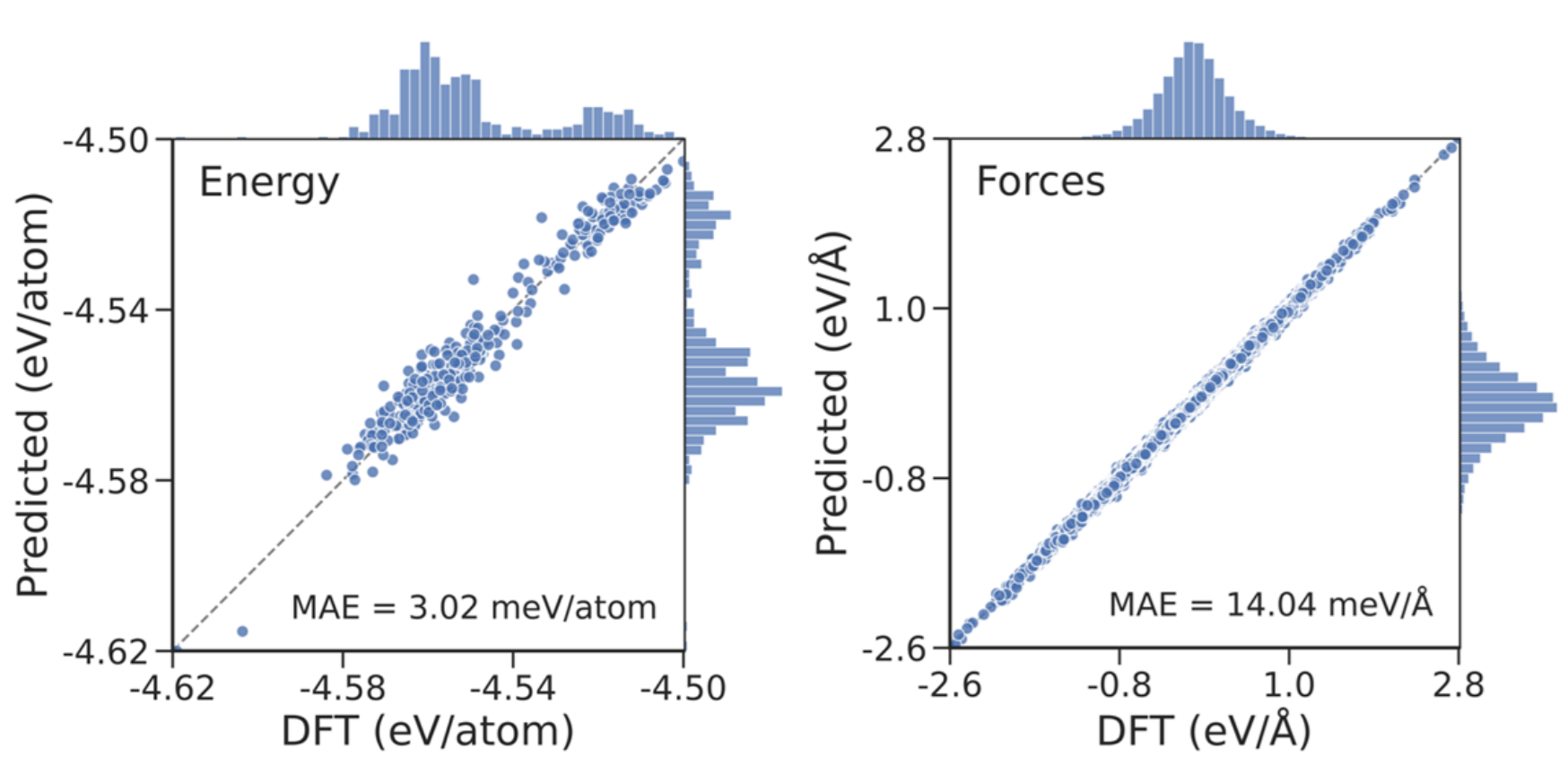}
  \caption{\textbf{Comparison between DFT (PBEsol+MBD) and ACE model predictions for energy and forces on the test set.}
    The ACE model was trained on data generated with the PBEsol with MBD functional.}
  \label{fig:comparison:of:dft:and:ace:potential:mbd}
\end{figure}

\begin{figure}[H]
  \centering
  \includegraphics[width=0.5\columnwidth]{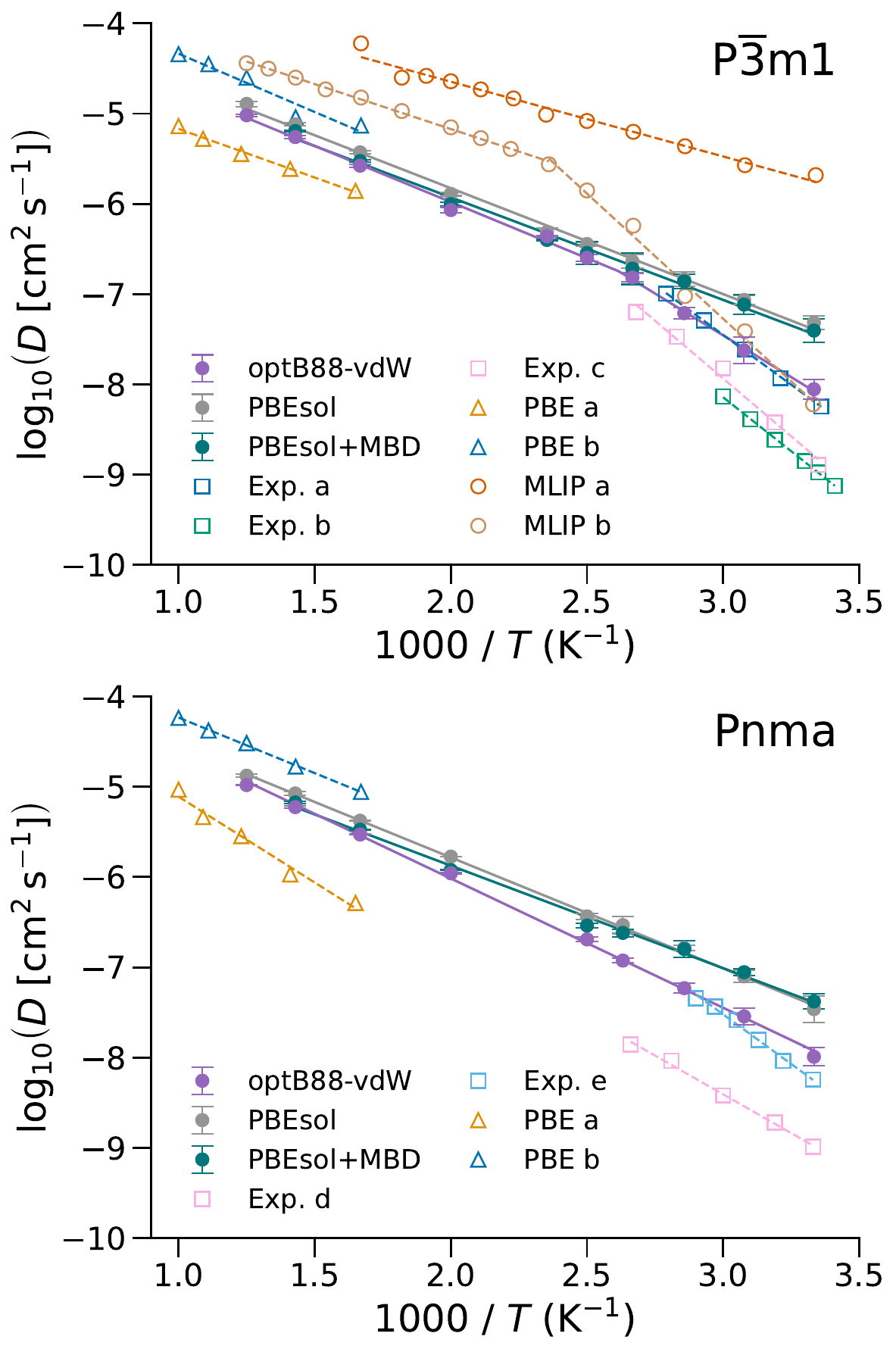}
  \caption{\textbf{Arrhenius plot of the \ce{Li+} diffusivity $D$ for the three exchange--correlation functionals tested}.
    Filled symbols represent results from this work, using ACE models trained on optB88-vdW, PBEsol, and PBEsol+MBD DFT data.
    Open symbols represent literature data.
    Experimental values: a~\cite{asano2018solid},  b~\cite{schlem2020mechanochemical}, c~\cite{chen2024accelerating}, d~\cite{hu2023revealing} and e~\cite{qiu2025scalable};
    AIMD simulations (PBE a~\cite{hu2023revealing} and PBE b~\cite{park2020theoretical});
    classical MD simulations using MLIPs (MLIP a, fitted to PBE DFT data)~\cite{qi2021bridging} and (MLIP b, fitted to  optB88-vdW DFT data)~\cite{qi2021bridging}.
    Error bars represent standard deviations from three independent MD runs.
  }
  \label{fig:sigma:vs:T:MBD}
\end{figure}

The PBEsol and PBEsol+MBD models give closely similar transport predictions (\fref{fig:sigma:vs:T:MBD} and \tref{tab:sigma:Ea:secondary}).
Both remain approximately linear in the Arrhenius representation over the sampled range, with fitted single-regime activation energies of 0.23--0.24~eV, and neither shows the change in Arrhenius slope obtained with optB88-vdW.
Both also overestimate the room-temperature conductivity: they give 3.4--4.2~\si{\milli\siemens\per\centi\meter} for \pmtmo, roughly an order of magnitude above the measured 0.10--0.51~\si{\milli\siemens\per\centi\meter}~\cite{asano2018solid,schlem2020mechanochemical,chen2024accelerating}.
By comparison, the optB88-vdW model gives a low-temperature activation energy of 0.37~eV for unstrained \pmtmo, within the experimental range of 0.40--0.51~eV, and a room-temperature conductivity of 0.77~\si{\milli\siemens\per\centi\meter} (Table~III of the main text).
Their low-temperature transport predictions are therefore less consistent with the experimental values than those obtained with optB88-vdW.
Thus, adding the MBD dispersion treatment to PBEsol does not by itself produce the crossover predicted by optB88-vdW; the comparison demonstrates sensitivity to the complete exchange--correlation treatment rather than identifying dispersion interactions alone as the origin of the crossover.

\begin{table}[tbh!]
  \centering
  \caption{\textbf{Li ionic conductivity $\sigma$ and activation energy $E_a$ from the secondary PBEsol and PBEsol+MBD ACE models.}
    Values are Nernst--Einstein estimates derived from the Li diffusivity and are reported at 300~K, obtained with the same procedure used for Table~III of the main text, where the corresponding optB88-vdW and experimental values are listed.}
  \label{tab:sigma:Ea:secondary}
  \begin{tabular}{lcccc}
    \hline
    Phase & $\sigma$ & $E_a$ & $T$       & Method          \\
          & (mS/cm)  & (eV)  & (K)       &                 \\
    \hline
    \multirow{2}{*}{\pmtmo} & 4.21 & 0.23 & 300--800 & MD (PBEsol)     \\
                            & 3.44 & 0.23 & 300--700 & MD (PBEsol+MBD) \\
    \hline
    \multirow{2}{*}{\pnma}  & 2.96 & 0.24 & 300--800 & MD (PBEsol)     \\
                            & 3.60 & 0.23 & 300--700 & MD (PBEsol+MBD) \\
    \hline
  \end{tabular}
\end{table}

\section{Migration-Energy Validation under Strain}
\label{sec:neb:validation}

We tested the final optB88-vdW ACE model on a representative direct Oct--Oct
migration event in the \pmtmo structure, using the 60-atom ordered realization
of the conventional cell employed for the DFT reference calculations.
The same symmetry orbit and directed edge were retained at fixed-cell isotropic
strains of $-2$\%, 0\%, and $+2$\% so that the strain dependence was evaluated
for an equivalent local migration environment.
For each strain state, the endpoints were relaxed with ACE at fixed cell, and a
nine-image band, including both endpoints, was initialized by image-dependent
pair-potential interpolation and optimized using the improved-tangent
climbing-image nudged elastic band (NEB) method as implemented in ASE.
The final climbing-image stage used a spring constant of
0.10~eV~\AA$^{-2}$ and a projected-force threshold of
0.03~eV~\AA$^{-1}$.

Every ACE-optimized image was then evaluated independently by a static DFT
calculation using VASP and the optB88-vdW settings used for the training data.
The calculations used a 520~eV plane-wave cutoff, a $2\times2\times2$
$\Gamma$-centered $k$-point mesh, an electronic convergence criterion of
$10^{-5}$~eV.
No DFT relaxation was performed for the individual images.

\begin{figure}[tbh!]
  \centering
  \includegraphics[width=0.95\linewidth]{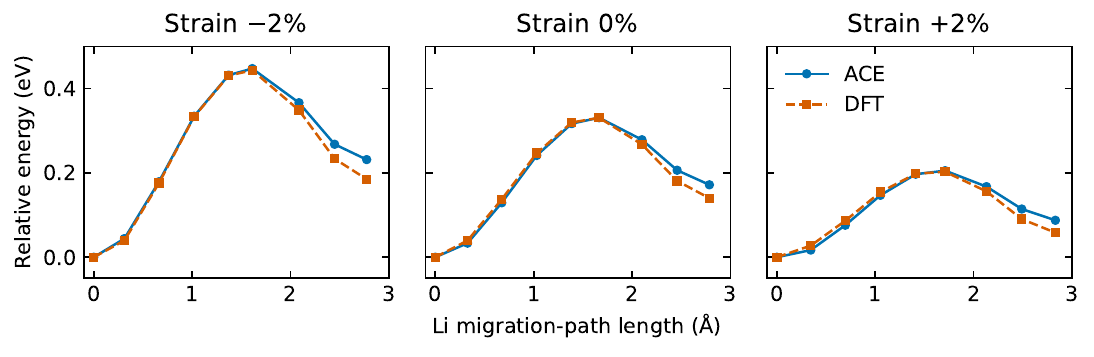}
  \caption{\textbf{DFT validation of the ACE Oct--Oct migration-energy profile under strain.}
    A representative Oct--Oct pathway in \pmtmo \lyc was optimized using the final ACE model at fixed-cell strains of $-2$\%, 0\%, and $+2$\%.
    Each of the nine ACE-optimized images was then evaluated independently using single-point DFT.
    For each method and strain state, the energy is reported relative to the leftmost image.
  }
  \label{fig:neb:validation}
\end{figure}

\fref{fig:neb:validation} shows the migration-energy profiles under strain.
\tref{tab:neb:validation} reports the quantitative comparison.
For each method and strain state, the energy zero was set independently to the
left endpoint.
The forward barrier is therefore
$\Delta E^\ddagger=\max_i(E_i-E_0)$, and the endpoint-energy difference is
$\Delta E_{B-A}=E_8-E_0$ for the directed event $A\rightarrow B$.
Because the two endpoints are inequivalent in the selected realization, the
reverse barrier, $\Delta E^\ddagger_{B\rightarrow A}=\max_i(E_i-E_8)$,
differs from the forward barrier.

\begin{table}[tbh!]
  \centering
  \caption{\textbf{ACE and DFT migration energetics for the representative
      Oct--Oct pathway under strain.}
    All quantities are in eV.
    $\Delta\Delta E^\ddagger$ is the corresponding barrier change relative to
    the unstrained structure; negative values denote barrier reduction.
    Forward and reverse barriers are both listed because the selected endpoints
    are inequivalent.
    The DFT values are single-point energies evaluated on the ACE-optimized
    images rather than results of an independently optimized DFT NEB.}
  \label{tab:neb:validation}
  \begin{tabular}{crrrrrrrrrr}
    \hline
    & \multicolumn{2}{c}{$\Delta E^\ddagger_{A\rightarrow B}$}
    & \multicolumn{2}{c}{$\Delta E^\ddagger_{B\rightarrow A}$}
    & \multicolumn{2}{c}{$\Delta E_{B-A}$}
    & \multicolumn{2}{c}{$\Delta\Delta E^\ddagger_{A\rightarrow B}$}
    & \multicolumn{2}{c}{$\Delta\Delta E^\ddagger_{B\rightarrow A}$} \\
    Strain & ACE & DFT & ACE & DFT & ACE & DFT & ACE & DFT & ACE & DFT \\
    \hline
    $-2$\% & 0.448 & 0.444 & 0.216 & 0.258 & 0.232 & 0.185 &  0.117 &  0.113 &  0.057 &  0.067 \\
    0\%    & 0.331 & 0.331 & 0.159 & 0.191 & 0.172 & 0.140 &  0.000 &  0.000 &  0.000 &  0.000 \\
    $+2$\% & 0.205 & 0.203 & 0.118 & 0.144 & 0.088 & 0.058 & $-0.126$ & $-0.128$ & $-0.041$ & $-0.047$ \\
    \hline
  \end{tabular}
\end{table}

The ACE and DFT forward barriers differ by at most 4~meV.
From $-2$\% to $+2$\% strain, they decrease by 0.242 and 0.241~eV,
respectively, showing that the final ACE model reproduces both the profile and
its strain dependence along this representative pathway.
The reverse barriers and the endpoint energy differences, however, differ by 26
to 47~meV.
These errors are smaller than the 0.24~eV decrease under strain, so the strain
trend is reliable, although differences in the barrier below about 50~meV are
within the error of this comparison, which can affect how accurately fine
changes in conductivity are resolved.
This comparison validates the DFT energies along the ACE pathway but does not
constitute an independently optimized DFT NEB or establish the statistics of
all migration events.

\section{Trajectory-Based \ce{Li+} Transport Analysis}

\subsection{Transport Coefficients and Arrhenius Analysis}
\label{sec:ionic:cond:calc}

\begin{figure}[tbh!]
  \centering
  \includegraphics[width=\linewidth]{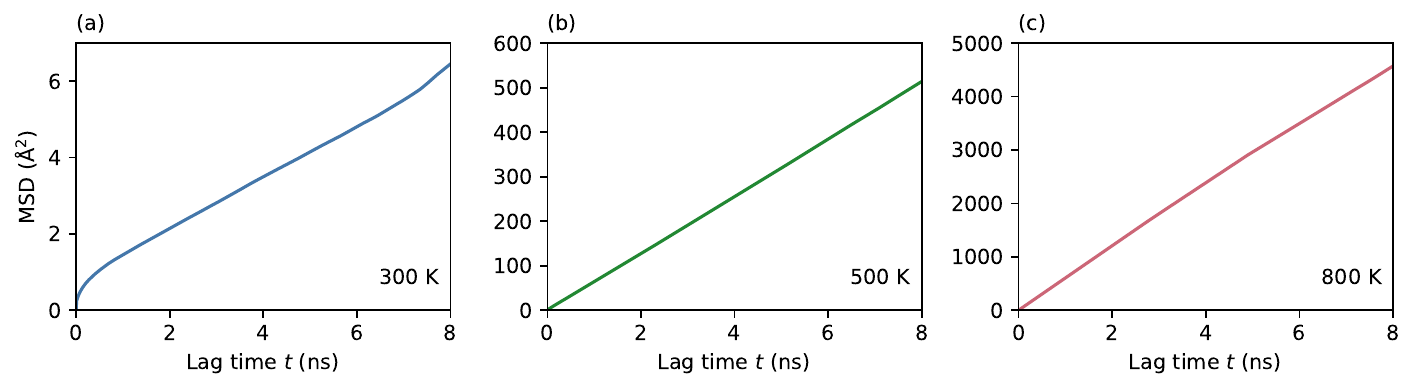}
  \caption{\textbf{Li mean squared displacement for unstrained \pnma \ce{Li3YCl6}.}
    MSD against the lag time $t$ at (a) 300~K, (b) 500~K, and (c) 800~K.
    The lag time is the separation between two configurations, not the elapsed simulation time: each point averages over every pair of configurations in the 10~ns production trajectory separated by $t$.
    The number of such pairs falls as $t$ approaches the trajectory length, so the curves are shown only up to $t=8$~ns, beyond which too few independent origins remain for a meaningful average.
    Note the different vertical scales; the MSD grows by roughly three orders of magnitude between 300 and 800~K.
    The diffusivity is obtained from the slope over the 200--800~ps lag window, where the MSD is linear in $t$.}
  \label{fig:msd:curves}
\end{figure}

The \ce{Li+} diffusivity $D$ is extracted from MD trajectories using the Einstein relation~\cite{einstein1905molekularkinetischen}, which connects the mean squared displacement (MSD) of the mobile ions to $D$:
\begin{equation}
  D = \frac{1}{2dNt} \sum_{i=1}^N  \langle | \mathbf{r}_i(t) - \mathbf{r}_i(0) |^2 \rangle
\end{equation}
where $N$ is the number of mobile ions, $d$ is the system dimensionality (typically $d=3$), and $\mathbf{r}_i(t)$ is the position vector of ion $i$ at time $t$.
The angular brackets denote an average over all time origins, so $t$ is the lag time separating two configurations rather than the elapsed simulation time.
$D$ is obtained by linear regression of the MSD against $t$ within the diffusive regime, where the slope is proportional to $2dD$; representative MSD curves are shown in \fref{fig:msd:curves}.
The first 1~ns of each simulation was discarded as equilibration, and each state point was then sampled using three independent 10~ns production trajectories, with configurations saved every 1~ps.
The stored production trajectories were analyzed in full, and the standard deviation of $\log_{10} D$ among the three independent trajectories is reported as the uncertainty.

\begin{figure}[tbh!]
  \centering
  \includegraphics[width=0.5\linewidth]{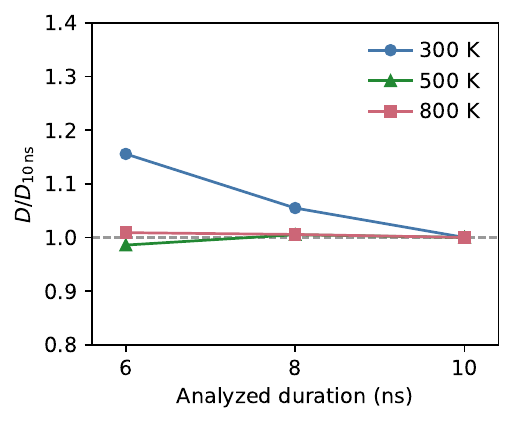}
  \caption{\textbf{Trajectory-length convergence of the Li diffusivity for unstrained \pnma \ce{Li3YCl6}.}
    Diffusivities calculated from cumulative 6, 8, and 10~ns portions of each trajectory, normalized by the corresponding 10~ns value.
    A ratio of unity, marked by the dashed line, indicates that the shorter analysis window already reproduces the 10~ns result.}
  \label{fig:msd:convergence}
\end{figure}

To assess sensitivity to trajectory length, we repeated the analysis for cumulative 6, 8, and 10~ns portions of representative unstrained \pnma trajectories at 300, 500, and 800~K (\fref{fig:msd:convergence}).
At 500 and 800~K, the diffusivities obtained from 6~ns differ from the corresponding 10~ns results by only 1.4\% and 0.9\%, respectively, showing that these estimates are already stable at 6~ns.
At 300~K, the normalized diffusivity decreases progressively from 1.156 at 6~ns to 1.055 at 8~ns and 1.000 at 10~ns.
This trend indicates convergence with increasing trajectory length, but it does not establish that the 10~ns result is fully converged.
Nevertheless, the 8 and 10~ns estimates differ by only 5.5\%, which supports the adequacy of the 10~ns trajectory while retaining the greater finite-sampling uncertainty at 300~K.
This duration-sensitivity check does not replace the independent-run uncertainty reported for each state point.
Because 300~K lies on the low-temperature branch of the two-region fits, we also considered how this residual sampling uncertainty propagates to the fitted crossover temperature.
The 5.5\% difference between the 8 and 10~ns estimates corresponds to a shift of 0.054 in $\ln D$, which is roughly two orders of magnitude smaller than the variation of $\ln D$ across the fitted temperature range.
Consistent with this, the temperature-grid sensitivity test described below leaves the fitted breakpoints essentially unchanged (\tref{tab:breakpoint:sensitivity}), indicating that they are not controlled by any single low-temperature point.

The ionic conductivity $\sigma$ is then derived from $D$ via the Nernst--Einstein equation~\cite{nernst1888kinetik}:
\begin{equation}
  \sigma = \frac{N}{V} \frac{q^2}{k_\text{B} T} D
\end{equation}
where $q$ is the charge of the carrier, $V$ is the volume of the simulation cell, $k_\text{B}$ is the Boltzmann constant, and $T$ is the temperature.
This estimate, denoted by $\sigma$ throughout the main text, is obtained from the tracer diffusivity and assumes that distinct ion--ion correlations are negligible.

\subsubsection{Collective Transport Correction}

The diffusivity $D$ reported in the main text is the tracer (self-)diffusivity obtained from individual Li displacements.
Its Nernst--Einstein conductivity estimate neglects distinct ion--ion correlations and is therefore not generally equal to the collective charge conductivity.
To assess distinct Li--Li correlations, we additionally calculated the collective diffusivity $D_\mathrm{coll}$ from the mean-squared total Li displacement,
\begin{equation}
  D_\mathrm{coll}
  =\lim_{t\rightarrow\infty}\frac{1}{6Nt}
  \left\langle\left|\sum_{i=1}^{N}\Delta\mathbf{r}_i(t)\right|^2\right\rangle.
\end{equation}
The corresponding collective conductivity is
\begin{equation}
  \sigma_\mathrm{coll}=\frac{N}{V}\frac{q^2}{k_\mathrm{B}T}D_\mathrm{coll}.
\end{equation}
The conventional Haven ratio is therefore
\begin{equation}
  H_\mathrm{R}=\frac{D}{D_\mathrm{coll}}
  =\frac{\sigma}{\sigma_\mathrm{coll}}.
\end{equation}
Here, $H_\mathrm{R}=1$ indicates no net correlation correction, whereas $H_\mathrm{R}<1$ indicates that positive distinct-ion correlations enhance net charge transport.

We also evaluated the collective MSDs at 300~K.
Although the tracer diffusivities at 300~K are supported by the independent trajectory-length test in \fref{fig:msd:convergence}, the collective MSDs did not yield stable, common diffusive slopes across the phase and strain states.
We therefore do not assign quantitative collective diffusivities or Haven ratios at 300~K.

\tref{tab:collective:transport} reports the quantitative analysis for both phases at 500 and 800~K under $-2$\%, 0\%, and $+2$\% strain.
For a consistent comparison, the tracer and collective MSDs were calculated from the same stored trajectories, averaged over all time origins, and fitted over the same lag-time window of 200--800~ps.

\begin{table*}[tbh!]
  \centering
  \caption{\textbf{Tracer and collective transport coefficients for \ce{Li3YCl6} at 500 and 800~K.}
    Diffusivities are in $10^{-6}$~\si{\centi\meter\squared\per\second}, and conductivities are in \si{\milli\siemens\per\centi\meter}.}
  \label{tab:collective:transport}
  \begin{tabular}{cccccccc}
    \hline
    Phase & $T$ (K) & Strain & $D$ & $D_\mathrm{coll}$ & $\sigma$ & $\sigma_\mathrm{coll}$ & $H_\mathrm{R}$ \\
    \hline
    \pmtmo & 500 & $-2$\% & 0.0394 & 0.0440 & 2.23  & 2.49  & 0.90 \\
            &     & 0\%    & 0.709  & 1.404  & 37.8  & 74.8  & 0.51 \\
            &     & $+2$\% & 3.181  & 6.636  & 160   & 333   & 0.48 \\
            & 800 & $-2$\% & 3.76    & 13.43   & 132   & 474   & 0.28 \\
            &     & 0\%    & 9.20    & 15.30   & 306   & 509   & 0.60 \\
            &     & $+2$\% & 18.26   & 21.55   & 573   & 676   & 0.85 \\
    \pnma   & 500 & $-2$\% & 0.218   & 0.512   & 12.2  & 28.7  & 0.43 \\
            &     & 0\%    & 1.056   & 1.876   & 55.7  & 99.0  & 0.56 \\
            &     & $+2$\% & 3.700   & 9.660   & 184   & 480   & 0.38 \\
            & 800 & $-2$\% & 4.22    & 8.45    & 148   & 296   & 0.50 \\
            &     & 0\%    & 10.09   & 16.71   & 333   & 551   & 0.60 \\
            &     & $+2$\% & 20.04   & 28.65   & 623   & 890   & 0.70 \\
    \hline
  \end{tabular}
\end{table*}

The collective estimates retain the ordering $\sigma_\mathrm{coll}(-2\%)<\sigma_\mathrm{coll}(0\%)<\sigma_\mathrm{coll}(+2\%)$ at both temperatures and in both phases, confirming that the strain trend is not an artifact of the Nernst--Einstein approximation.
For the common 200--800~ps window, $H_\mathrm{R}<1$ at all 12 state points, indicating that distinct-ion correlations enhance net charge transport over this intermediate-time range.
The strength of the correction is state dependent, with $H_\mathrm{R}$ ranging from 0.28 to 0.90.
Fits over 100--400, 400--1200, and 800--1600~ps show that several exact ratios remain sensitive to the lag-time window as the number of independent collective displacements decreases.
We therefore use this analysis to test the robustness of the qualitative strain conclusion rather than to assign a universal Haven ratio or replace the consistently reported Nernst--Einstein conductivities with apparently precise dc collective values.

\subsubsection{Arrhenius Fitting and Model Selection}

The Arrhenius analysis was performed using the arithmetic-mean diffusivity at each sampled temperature.
For each phase and strain state, we compared a single Arrhenius line with the continuous two-region model
\begin{equation}
  \ln D = a + b x + c\max(0,x-x_c),
  \qquad x=1/T, \qquad x_c=1/T_c,
\end{equation}
where the breakpoint $T_c$ was optimized freely while retaining at least three sampled temperatures on each side.
The single-line model has two fitted parameters and the two-region model has four, including $T_c$.
For a single-line fit, $D_0=\exp(a)$.
For a two-region fit, we report $D_0=\exp(a)$ from the high-temperature branch and characterize the low-temperature branch by $E_a$ and $D(T_c)=\exp(a+b x_c)$; its formal $1/T=0$ intercept lies outside the fitted domain and is therefore not reported.
The reported $D_0$ values are treated as extrapolated fit parameters without an independent microscopic interpretation.
For a two-region fit, $D_0$ is not an independent quantity: because the two branches meet at the breakpoint, the high-temperature intercept is fixed by the corresponding slope and by the breakpoint diffusivity through $D_0=D(T_c)\exp[E_a/(k_\text{B}T_c)]$.
Trends in $D_0$ across phase and strain states therefore largely reflect this constraint rather than independent microscopic information.
Model preference was evaluated using the small-sample corrected Akaike information criterion (AICc)~\cite{hurvich1989regression},
\begin{equation}
  \mathrm{AICc}
  = n\ln\!\left(\frac{\mathrm{RSS}}{n}\right)
    +2k+\frac{2k(k+1)}{n-k-1},
\end{equation}
where $n$ is the number of sampled temperatures, $k$ is the number of fitted parameters, and RSS is the residual sum of squares in $\ln D$.
We define $\Delta\mathrm{AICc}=\mathrm{AICc}_{\mathrm{single}}-\mathrm{AICc}_{\mathrm{two}}$, so positive values favor the two-region model.
We retain the two-region model only when $\Delta\mathrm{AICc}>4$ and use the single Arrhenius line otherwise.

\tref{tab:breakpoint:sensitivity} reports the all-temperature fits and a sensitivity test in which the sampled temperature nearest the previously used boundary was removed (425~K for \pmtmo and 380~K for \pnma).
The candidate $T_c$ is listed even when the single-line model was selected, to show why that breakpoint was not retained in the revised Arrhenius plot.

\begin{table}[H]
  \centering
  \small
  \caption{\textbf{Free-breakpoint model selection and temperature-grid sensitivity.}
    The selected model is used in the revised strain-dependent Arrhenius plot.
    $T_c^{(2)}$ is the candidate breakpoint from the continuous two-region fit.
    The final two columns give the corresponding quantities after removing 425~K for \pmtmo or 380~K for \pnma.}
  \label{tab:breakpoint:sensitivity}
  \begin{tabular}{llcrrrr}
    \hline
    Phase & Strain & Selected model & $T_c^{(2)}$ (K) & $\Delta\mathrm{AICc}$ & $T_{c,\mathrm{rm}}^{(2)}$ (K) & $\Delta\mathrm{AICc}_{\mathrm{rm}}$ \\
    \hline
    \multirow{5}{*}{\pmtmo}
      & $-2$\% & two    & 508 &  7.2 & 508 &  2.8 \\
      & $-1$\% & single & 507 & $-2.8$ & 524 & $-6.3$ \\
      & 0\%    & two    & 382 &  9.2 & 375 &  7.0 \\
      & $+1$\% & two    & 393 & 21.8 & 393 & 15.5 \\
      & $+2$\% & two    & 415 & 35.0 & 418 & 26.9 \\
    \hline
    \multirow{5}{*}{\pnma}
      & $-2$\% & two    & 432 & 11.2 & 441 &  6.3 \\
      & $-1$\% & single & 393 &  2.0 & 400 & $-1.8$ \\
      & 0\%    & single & 473 &  1.7 & 451 &  1.7 \\
      & $+1$\% & two    & 407 &  6.6 & 410 &  0.2 \\
      & $+2$\% & two    & 355 & 21.4 & 356 & 13.4 \\
    \hline
  \end{tabular}
\end{table}

The data therefore do not support a strain-invariant crossover temperature.
For \pmtmo at $-1$\% strain and \pnma at $-1$\% and 0\% strain, the available temperature grid does not reliably distinguish two Arrhenius regions, and the simpler single-line model is used.
For the retained two-region cases, removing the temperature nearest the former boundary changes the fitted $T_c$ by at most 9~K, although the AICc preference weakens for some of the more sparsely resolved cases.
Because $E_a$ and the fitted intercept $D_0$ are obtained from the same regression, their uncertainties are correlated and both depend on the fitted temperature range; we therefore report $D_0$ as a fit parameter and do not interpret its variation across phase and strain states.

\begin{table}[H]
  \centering
  \small
  \caption{\textbf{Diffusivity prefactors from the Arrhenius models selected for the strain-dependent analysis.}
    The values are in \si{\centi\meter\squared\per\second} and were obtained using all temperatures shown in Figure~8 of the main text.
    For a two-region model, $D_0$ is the high-temperature intercept and $D(T_c)$ is the breakpoint diffusivity; for a single-line model, $D_0$ is the line intercept.}
  \label{tab:arrhenius:prefactors}
  \begin{tabular}{llcccc}
    \hline
    Phase & Strain & Model & $T_c$ (K) & $D(T_c)$ & $D_0$ \\
    \hline
    \multirow{5}{*}{\pmtmo}
      & $-2$\% & two    & 508 & $2.34\times10^{-7}$ & $5.78\times10^{-4}$ \\
      & $-1$\% & single & -   & -                      & $3.26\times10^{-4}$ \\
      & 0\%    & two    & 382 & $1.79\times10^{-7}$ & $3.22\times10^{-4}$ \\
      & $+1$\% & two    & 393 & $4.05\times10^{-7}$ & $4.43\times10^{-4}$ \\
      & $+2$\% & two    & 415 & $1.31\times10^{-6}$ & $3.57\times10^{-4}$ \\
    \hline
    \multirow{5}{*}{\pnma}
      & $-2$\% & two    & 432 & $7.63\times10^{-8}$ & $4.93\times10^{-4}$ \\
      & $-1$\% & single & -   & -                      & $3.30\times10^{-4}$ \\
      & 0\%    & single & -   & -                      & $7.06\times10^{-4}$ \\
      & $+1$\% & two    & 407 & $6.71\times10^{-7}$ & $3.56\times10^{-4}$ \\
      & $+2$\% & two    & 355 & $6.30\times10^{-7}$ & $3.27\times10^{-4}$ \\
    \hline
  \end{tabular}
\end{table}

\subsubsection{Provenance of the Experimental Conductivity Data}

The experimental data compared with the simulations in the main text were all
obtained by electrochemical impedance spectroscopy on powder-derived
polycrystalline pellets, rather than on single crystals or thin
films~\cite{asano2018solid,schlem2020mechanochemical,chen2024accelerating,hu2023revealing,qiu2025scalable}.
One of these studies resolved a single conductive response with no additional
grain-boundary contribution~\cite{asano2018solid}, and one reported pellets of
geometric density above 85\% whose response was considered predominantly
bulk-like~\cite{schlem2020mechanochemical}; the remaining studies did not
separate bulk from grain-boundary contributions.
These values are therefore treated as total pellet conductivities rather than
independently resolved bulk conductivities.

\subsection{Probability Density and Path Free Energy}

\begin{figure}[H]
  \centering
  \includegraphics[width=1\textwidth]{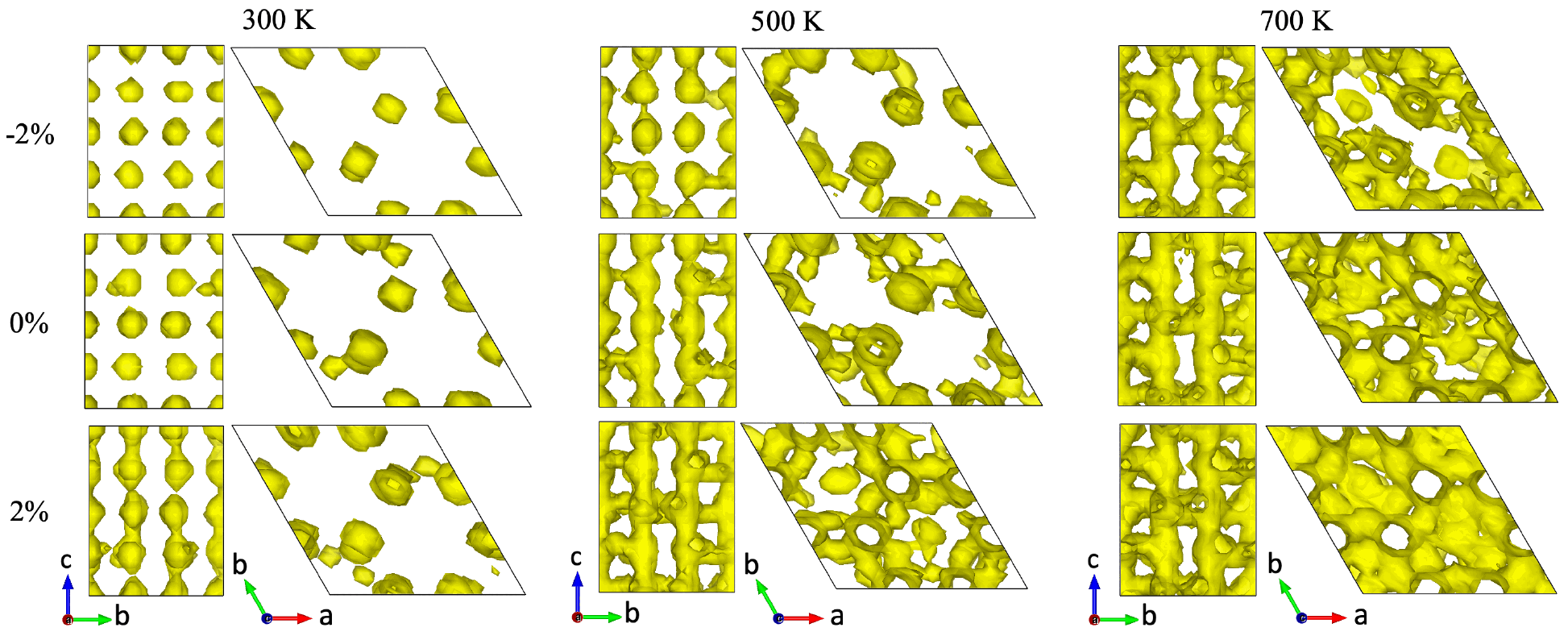}
  \caption{\textbf{\ce{Li+} probability density in \lyc (\pmtmo) under various strain conditions.}
  }
  \label{fig:density:p3m1}
\end{figure}

\begin{figure}[H]
  \centering
  \includegraphics[width=1\textwidth]{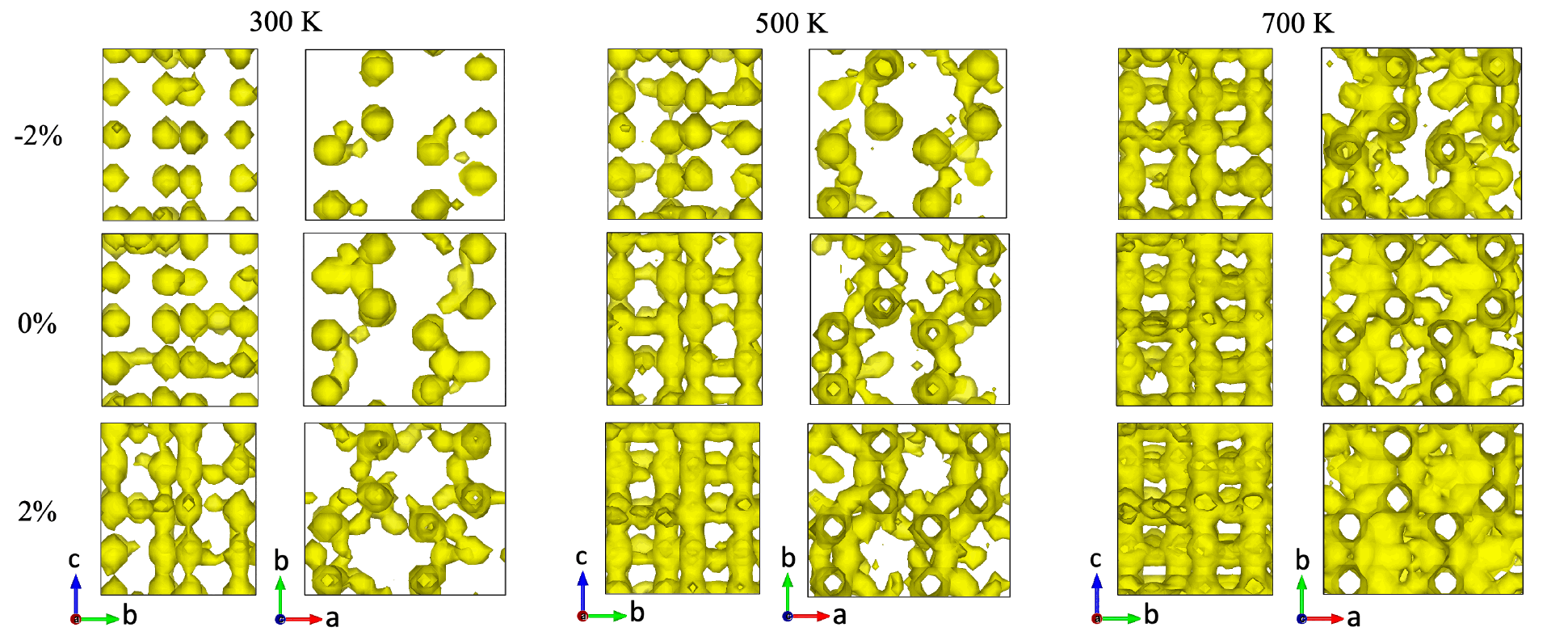}
  \caption{\textbf{\ce{Li+} probability density in \lyc (\pnma) under various strain conditions.}
  }
  \label{fig:density:pnma}
\end{figure}

The \ce{Li+} probability density maps shown in the main text were computed from 500~ps MD simulations
employing the \texttt{pymatgen-analysis-diffusion} package~\cite{Deng2016,Ong2013} to map the trajectory-derived ionic density onto a real-space grid.
The probability density $\rho(\mathbf{r})$ reflects the time-averaged likelihood
of finding a \ce{Li+} ion at a given position $\mathbf{r}$ in the simulation cell,
constructed by accumulating ionic positions across all MD snapshots onto a uniform
grid with a spatial interval of 0.5~\AA.
Path-resolved free-energy profiles were obtained separately, from the full
production trajectories rather than from the density maps above.
Each migration path connecting two neighboring sites of the relaxed framework
is treated as a tube: the \ce{Li+} positions of every stored frame that lie
within 1.25~\AA\ of the path axis are projected onto it, and the projected
coordinate is accumulated in 100 equal bins spanning the path, with a Gaussian
smoothing of 0.03~\AA\ applied along $s$.
Positions are referred to the \ce{Cl-}/\ce{Y^3+} framework of each frame, so
that global drift of the cell does not broaden the distribution.
Symmetry-equivalent paths are pooled, so that the resulting $P(s)$ is the
fraction of sampled \ce{Li+} positions found at path coordinate $s$ within that
path family.
The free energy along the path then follows by Boltzmann
inversion~\cite{roux1995calculation},
\begin{equation}
  \Delta F(s) = -k_\mathrm{B}T \ln \frac{P(s)}{P_{\max}},
  \label{eq:boltzmann:inversion}
\end{equation}
where $P_{\max}$ places the free-energy zero at the most probable point of the
path.
The forward barrier $\Delta F^\ddagger$ is the difference between the maximum of
$\Delta F(s)$ and the minimum in the starting basin.
The pooled forward barriers obtained from these profiles are summarized in
\tref{tab:path:free-energy} for both phases at three temperatures and three
strain states.
Every state was analyzed with identical settings: all stored frames of the
production trajectory were used, the first 10\% of each trajectory was
discarded as equilibration, and the quoted uncertainties are standard
deviations from five contiguous blocks of the remainder.
Coarser frame sampling was found to bias the barriers upward, because the
transition-region bins that set the peak of $\Delta F(s)$ are the most sparsely
populated, so no frames were skipped within the analyzed interval.
Regions of high $\rho(\mathbf{r})$ correspond to preferred \ce{Li+} residence sites
and low-barrier migration pathways, providing an intuitive visualization of the
diffusion landscape.
The isosurfaces shown in the main text correspond to a level of
$5\times 10^{-5}$~e/Bohr$^{3}$, chosen to highlight the interconnected
network of \ce{Li+} conduction channels while suppressing low-occupancy regions.
The probability density maps are provided in \fref{fig:density:p3m1} and \fref{fig:density:pnma} for the \pmtmo and \pnma phases, respectively.

\begin{table*}[tbh!]
  \centering
  \caption{\textbf{Trajectory-derived path free-energy barriers.}
    $\Delta F^\ddagger$ is the pooled forward barrier of each path family, and the uncertainty is the standard deviation across five contiguous trajectory blocks.
    The last two columns give the number of individual migration channels for which a finite barrier could be resolved, out of the total enumerated for that path family, and therefore indicate how completely the trajectory samples each family.
    $^{a}$This state resolves only half of its Oct--Oct channels and its trajectory blocks are degenerate, so no block standard deviation is quoted and the state is excluded from the discussion.}
  \label{tab:path:free-energy}
  \begin{tabular}{ccccccc}
    \hline
    & & & \multicolumn{2}{c}{$\Delta F^\ddagger$ (eV)} & \multicolumn{2}{c}{Resolved channels} \\
    \cline{4-5}\cline{6-7}
    Phase & $T$ (K) & Strain (\%) & Oct--Oct & Oct--Tet--Oct & Oct--Oct & Oct--Tet--Oct \\
    \hline
    \multirow{9}{*}{\pmtmo}
      & 300 & $-2$$^{a}$ & $0.555$ & $0.607$ & 334/648 & 1002/3888 \\
      & 300 & $0$ & $0.250\pm0.008$ & $0.308\pm0.121$ & 644/648 & 1914/3888 \\
      & 300 & $+2$ & $0.183\pm0.003$ & $0.271\pm0.022$ & 648/648 & 1950/3888 \\
      & 500 & $-2$ & $0.353\pm0.008$ & $0.414\pm0.024$ & 648/648 & 2213/3888 \\
      & 500 & $0$ & $0.267\pm0.001$ & $0.318\pm0.006$ & 648/648 & 2676/3888 \\
      & 500 & $+2$ & $0.210\pm0.001$ & $0.266\pm0.001$ & 648/648 & 2823/3888 \\
      & 800 & $-2$ & $0.376\pm0.003$ & $0.473\pm0.009$ & 648/648 & 2840/3888 \\
      & 800 & $0$ & $0.317\pm0.001$ & $0.410\pm0.005$ & 648/648 & 2815/3888 \\
      & 800 & $+2$ & $0.269\pm0.001$ & $0.350\pm0.006$ & 648/648 & 2849/3888 \\
    \hline
    \multirow{9}{*}{\pnma}
      & 300 & $-2$ & $0.303\pm0.058$ & $0.275\pm0.016$ & 730/864 & 2423/5184 \\
      & 300 & $0$ & $0.236\pm0.004$ & $0.249\pm0.016$ & 864/864 & 2622/5184 \\
      & 300 & $+2$ & $0.169\pm0.002$ & $0.226\pm0.012$ & 864/864 & 2617/5184 \\
      & 500 & $-2$ & $0.343\pm0.003$ & $0.326\pm0.006$ & 864/864 & 3549/5184 \\
      & 500 & $0$ & $0.271\pm0.003$ & $0.272\pm0.003$ & 864/864 & 3715/5184 \\
      & 500 & $+2$ & $0.212\pm0.001$ & $0.229\pm0.001$ & 864/864 & 3751/5184 \\
      & 800 & $-2$ & $0.373\pm0.003$ & $0.440\pm0.002$ & 864/864 & 3810/5184 \\
      & 800 & $0$ & $0.321\pm0.002$ & $0.383\pm0.002$ & 864/864 & 3765/5184 \\
      & 800 & $+2$ & $0.262\pm0.001$ & $0.328\pm0.002$ & 864/864 & 3773/5184 \\
    \hline
  \end{tabular}
\end{table*}

In \pmtmo the Oct--Tet--Oct barrier exceeds the Oct--Oct barrier in every state
examined, consistent with lower transverse-pathway accessibility.
The difference falls from $2.2$ to $1.2\,k_\mathrm{B}T$ between 300 and 500~K at
zero strain and changes little above 500~K, so both path families become more
thermally accessible while retaining their ordering.
At 300~K the difference is largest under tension, $0.088$~eV at $+2$\%, so the
transverse pathway that tensile strain makes accessible remains the more
difficult of the two.
In \pnma the two families are much closer, and the ordering reverses under
compression at 300 and 500~K, where the Oct--Tet--Oct barrier falls below the
Oct--Oct barrier; the \pmtmo ordering is recovered at 800~K and under tension.
The Oct--Tet--Oct profiles are heterogeneous across local Li environments, and
a finite barrier is resolved for only about half to three quarters of the
enumerated Oct--Tet--Oct channels, the fraction increasing with temperature.
Accordingly, the pooled values quantify the probability distribution sampled by
the finite trajectories; they neither establish a finite barrier for every
local orbit nor determine pathway dominance by themselves.

\subsection{Diffusion Anisotropy}

\begin{figure}[H]
  \centering
  \includegraphics[width=.8\textwidth]{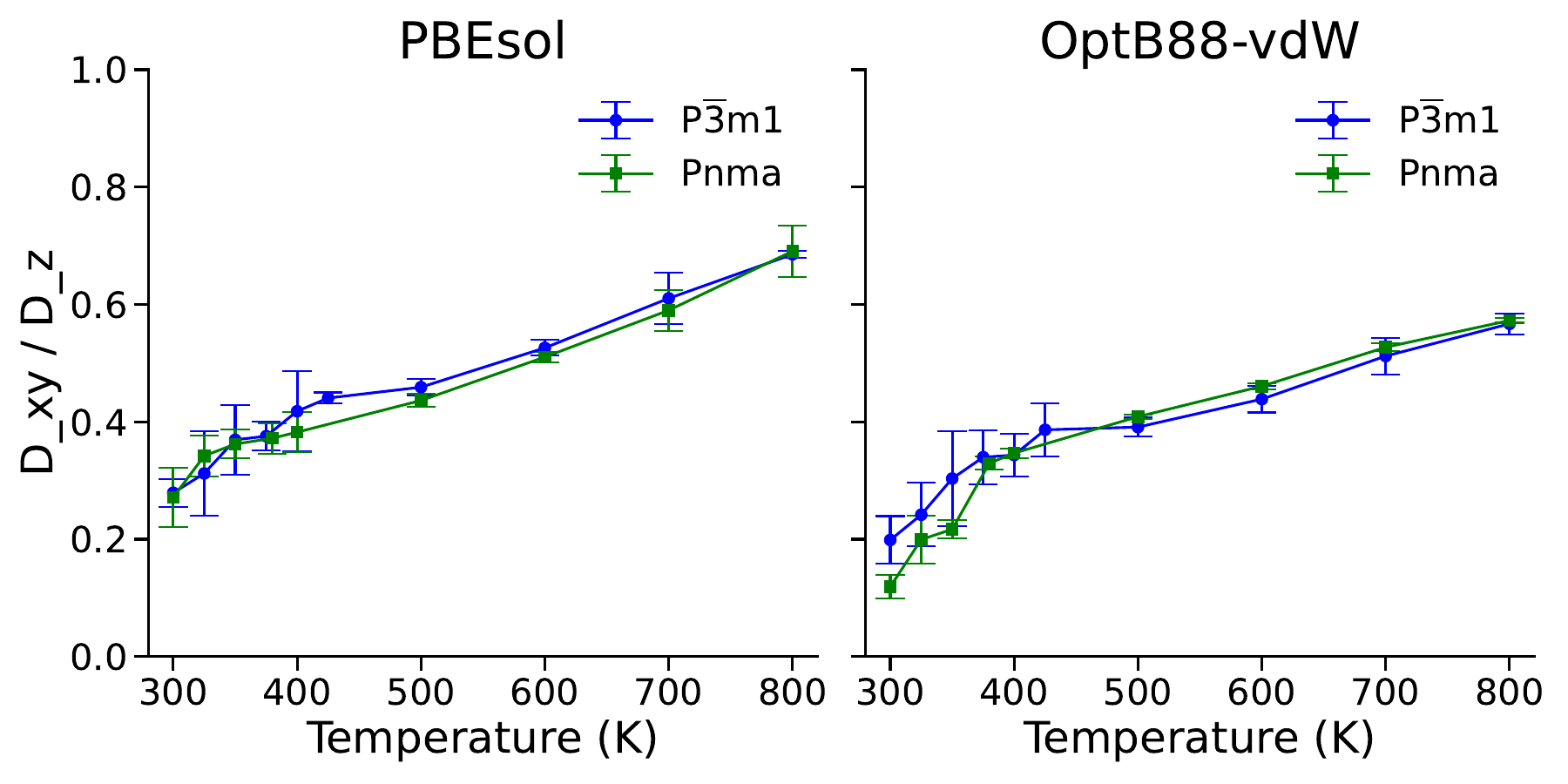}
  \caption{\textbf{Diffusivity ratio of \ce{Li+} along different directions.}}
  \label{fig:diffusivity:ratio:dft:functional}
\end{figure}

To characterize the anisotropy of \ce{Li+} diffusion, we plot the ratio
$D_{xy}/D_{z}$ for different DFT functionals in
\fref{fig:diffusivity:ratio:dft:functional}.
At low temperatures, \ce{Li+} migration is primarily restricted to the
$z$-axis via the Oct--Oct pathway, yielding $D_{xy}/D_{z} \ll 1$.
Above the critical temperature $T_c$, activation of the Oct--Tet--Oct
pathway drives $D_{xy}/D_{z}$ toward unity, signaling diffusion through a
connected three-dimensional pathway network.

\subsection{Site Occupancy}
\label{calculation:site}
To quantify the dynamic distribution of \ce{Li+} ions, we analyzed the local coordination environment based on the distance between each \ce{Li+} and its neighbors throughout the MD trajectories.
The coordination number ($CN$) of a \ce{Li+} ion is defined as the number of \ce{Cl-} neighbors within 3.5~\AA.
This fixed operational cutoff lies near the first minimum of the Li--Cl radial distribution function (RDF).

The sites are categorized into four groups based on $CN$: octahedral ($CN=6$), tetrahedral ($CN=4$), intermediate ($CN=5$), and others ($CN=3$, $7$--$9$).
The occupancy for each site type $k$ is defined as the fraction of \ce{Li+} ions found in that environment across all sampled frames:
\begin{equation}
  \text{Site Fraction}_k = \frac{\sum_{t=1}^{N_\text{frames}} N_{k,t}}{N_\text{total} \times N_\text{frames}} \times 100\%
\end{equation}
where $N_{k,t}$ is the number of \ce{Li+} ions in type $k$ at frame $t$, $N_\text{total}$ is the number of \ce{Li+} ions in the simulation cell, and $N_\text{frames}$ is the total number of MD frames analyzed.

\begin{figure}[tbh!]
  \centering
  \includegraphics[width=\textwidth]{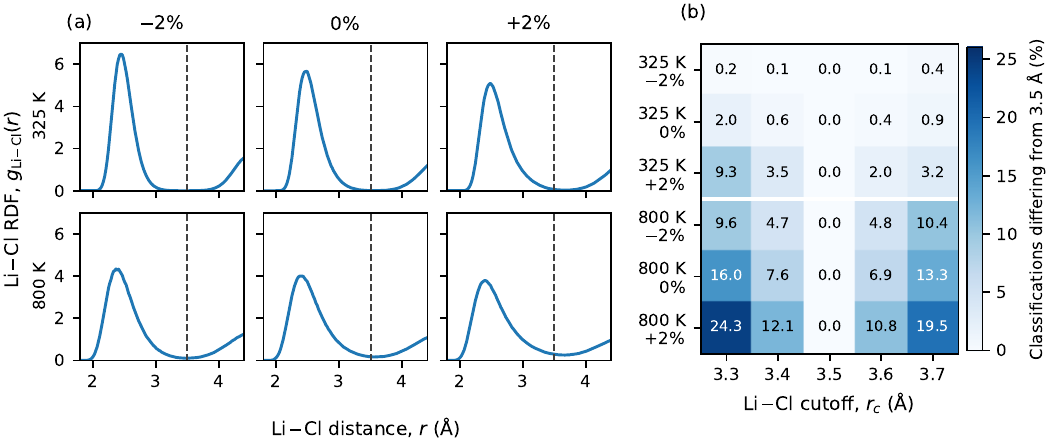}
  \caption{\textbf{Li coordination cutoff analysis for \pmtmo.}
    (a) Full-trajectory Li--Cl RDFs at $-2$\%, 0\%, and $+2$\% strain and 325 and 800~K; dashed lines mark the fixed 3.5~\AA\ cutoff.
    (b) Percentage of instantaneous coordination classifications that differ from the 3.5~\AA\ result when the cutoff is varied from 3.3 to 3.7~\AA.}
  \label{fig:cn:p3m1}
\end{figure}

\begin{figure}[tbh!]
  \centering
  \includegraphics[width=\textwidth]{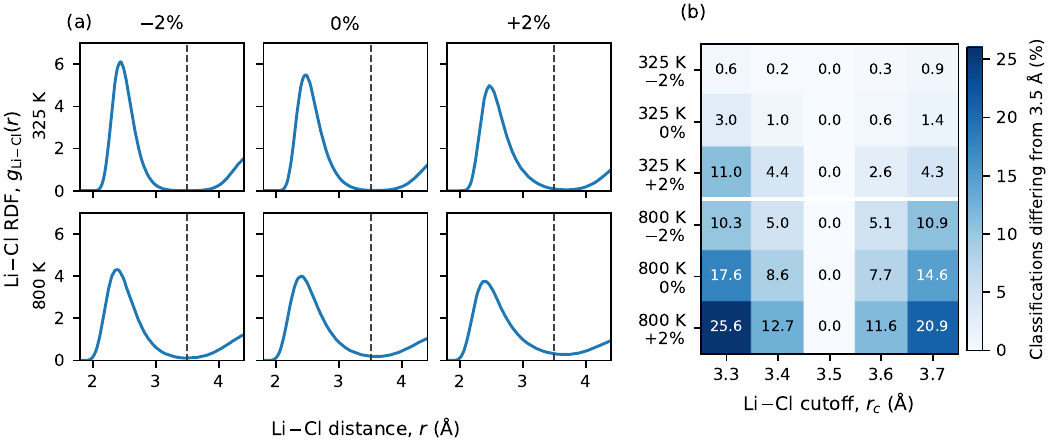}
  \caption{\textbf{Li coordination cutoff analysis for \pnma.}
    (a) Full-trajectory Li--Cl RDFs at $-2$\%, 0\%, and $+2$\% strain and 325 and 800~K; dashed lines mark the fixed 3.5~\AA\ cutoff.
    (b) Percentage of instantaneous coordination classifications that differ from the 3.5~\AA\ result when the cutoff is varied from 3.3 to 3.7~\AA.}
  \label{fig:cn:pnma}
\end{figure}

We tested the dynamic and cutoff robustness of this assignment using a matched set of 12 trajectories: both phases at $-2$\%, 0\%, and $+2$\% strain, each at 325 and 800~K.
Every tenth stored frame was analyzed, corresponding to a 10~ps sampling interval.
Each state contributed 1001 analyzed frames except the compressive 800~K trajectories, which contributed 480 frames for \pmtmo and 574 frames for \pnma.
The full-trajectory Li--Cl RDFs in \fref{fig:cn:p3m1} and \fref{fig:cn:pnma} show that 3.5~\AA\ lies in the low-probability region near the first-shell minimum for all selected states, although the minimum shifts to larger distance under tensile strain.

We additionally repeated the instantaneous site assignment with cutoffs from 3.3 to 3.7~\AA.
For each alternative cutoff, a classification is counted as different when the same Li configuration changes among the Oct, Tet, $CN=5$, and other categories relative to its classification at 3.5~\AA.
\tref{tab:cn:robustness} reports the fraction of classifications changed by the neighboring 3.4 and 3.6~\AA\ choices and the fraction having a Li--Cl distance within 0.10~\AA\ of the baseline cutoff.
The assignment is weakly cutoff-dependent at 325~K but becomes less precise in the thermally broadened tensile states at 800~K, where up to 22.4\% of configurations are cutoff-adjacent.
Accordingly, absolute instantaneous Oct and Tet fractions in these states should be interpreted as cutoff-dependent populations rather than unique structural labels.
Nevertheless, the principal strain trend is unchanged over the full 3.3--3.7~\AA\ interval.
For each cutoff $r_c$, we quantify this trend using the ratios
\begin{equation}
  R_{\mathrm{Tet}}^{\pm}(r_c)
  = \frac{f_{\mathrm{Tet}}(\pm2\%,r_c)}
  {f_{\mathrm{Tet}}(0\%,r_c)}.
\end{equation}
As shown in \tref{tab:cn:tet:ratios}, $R_{\mathrm{Tet}}^{-}<1$ and
$R_{\mathrm{Tet}}^{+}>1$ for every tested cutoff, phase, and temperature.
Thus, compression consistently decreases and tension consistently increases the
Tet population relative to the unstrained state.  The precise enhancement factor
is cutoff dependent, especially at 325~K where the unstrained Tet population is
very small, but the direction of the strain response is unchanged.

\begin{table}[tbh!]
  \centering
  \caption{\textbf{Cutoff sensitivity of the Li coordination assignment.}
    ``Adjacent'' is the fraction of Li configurations with a Li--Cl distance within $\pm0.10$~\AA\ of 3.5~\AA.
    The final column gives the fraction of classifications that differs from the 3.5~\AA\ result when the cutoff is replaced by 3.4 or 3.6~\AA.}
  \label{tab:cn:robustness}
  \begin{tabular}{cccrrr}
    \hline
    Phase & Strain & $T$ (K) & Frames & Adjacent (\%) & Changed at 3.4/3.6~\AA\ (\%) \\
    \hline
    \pmtmo & $-2$\% & 325 & 1001 & 0.16 & 0.07/0.10 \\
    \pmtmo & 0\%    & 325 & 1001 & 0.96 & 0.63/0.36 \\
    \pmtmo & $+2$\% & 325 & 1001 & 5.25 & 3.47/1.98 \\
    \pmtmo & $-2$\% & 800 &  480 & 8.88 & 4.69/4.82 \\
    \pmtmo & 0\%    & 800 & 1001 & 13.46 & 7.63/6.94 \\
    \pmtmo & $+2$\% & 800 & 1001 & 21.08 & 12.05/10.77 \\
    \pnma  & $-2$\% & 325 & 1001 & 0.46 & 0.20/0.28 \\
    \pnma  & 0\%    & 325 & 1001 & 1.53 & 0.97/0.61 \\
    \pnma  & $+2$\% & 325 & 1001 & 6.72 & 4.37/2.64 \\
    \pnma  & $-2$\% & 800 &  574 & 9.41 & 4.97/5.11 \\
    \pnma  & 0\%    & 800 & 1001 & 15.02 & 8.57/7.73 \\
    \pnma  & $+2$\% & 800 & 1001 & 22.45 & 12.74/11.56 \\
    \hline
  \end{tabular}
\end{table}

\begin{table}[tbh!]
  \centering
  \caption{\textbf{Cutoff dependence of the relative Tet population.}
    $R_{\mathrm{Tet}}^{-}$ and $R_{\mathrm{Tet}}^{+}$ are the Tet fractions at
    $-2$\% and $+2$\% strain, respectively, divided by the Tet fraction at zero
    strain using the same Li--Cl cutoff $r_c$.}
  \label{tab:cn:tet:ratios}
  \begin{tabular}{cccrr}
    \hline
    Phase & $T$ (K) & $r_c$ (\AA) & $R_{\mathrm{Tet}}^{-}$ & $R_{\mathrm{Tet}}^{+}$ \\
    \hline
    \pmtmo & 325 & 3.3 & 0.139 & 9.070 \\
            &     & 3.4 & 0.212 & 8.583 \\
            &     & 3.5 & 0.254 & 6.815 \\
            &     & 3.6 & 0.275 & 5.123 \\
            &     & 3.7 & 0.280 & 4.060 \\
            & 800 & 3.3 & 0.364 & 2.373 \\
            &     & 3.4 & 0.357 & 2.567 \\
            &     & 3.5 & 0.352 & 2.652 \\
            &     & 3.6 & 0.334 & 2.733 \\
            &     & 3.7 & 0.295 & 2.835 \\
    \pnma  & 325 & 3.3 & 0.362 & 4.821 \\
            &     & 3.4 & 0.414 & 4.338 \\
            &     & 3.5 & 0.432 & 3.950 \\
            &     & 3.6 & 0.424 & 3.738 \\
            &     & 3.7 & 0.386 & 3.739 \\
            & 800 & 3.3 & 0.362 & 2.179 \\
            &     & 3.4 & 0.349 & 2.312 \\
            &     & 3.5 & 0.344 & 2.426 \\
            &     & 3.6 & 0.335 & 2.519 \\
            &     & 3.7 & 0.303 & 2.649 \\
    \hline
  \end{tabular}
\end{table}

\begin{figure}[H]
  \centering
  \includegraphics[width=0.8\textwidth]{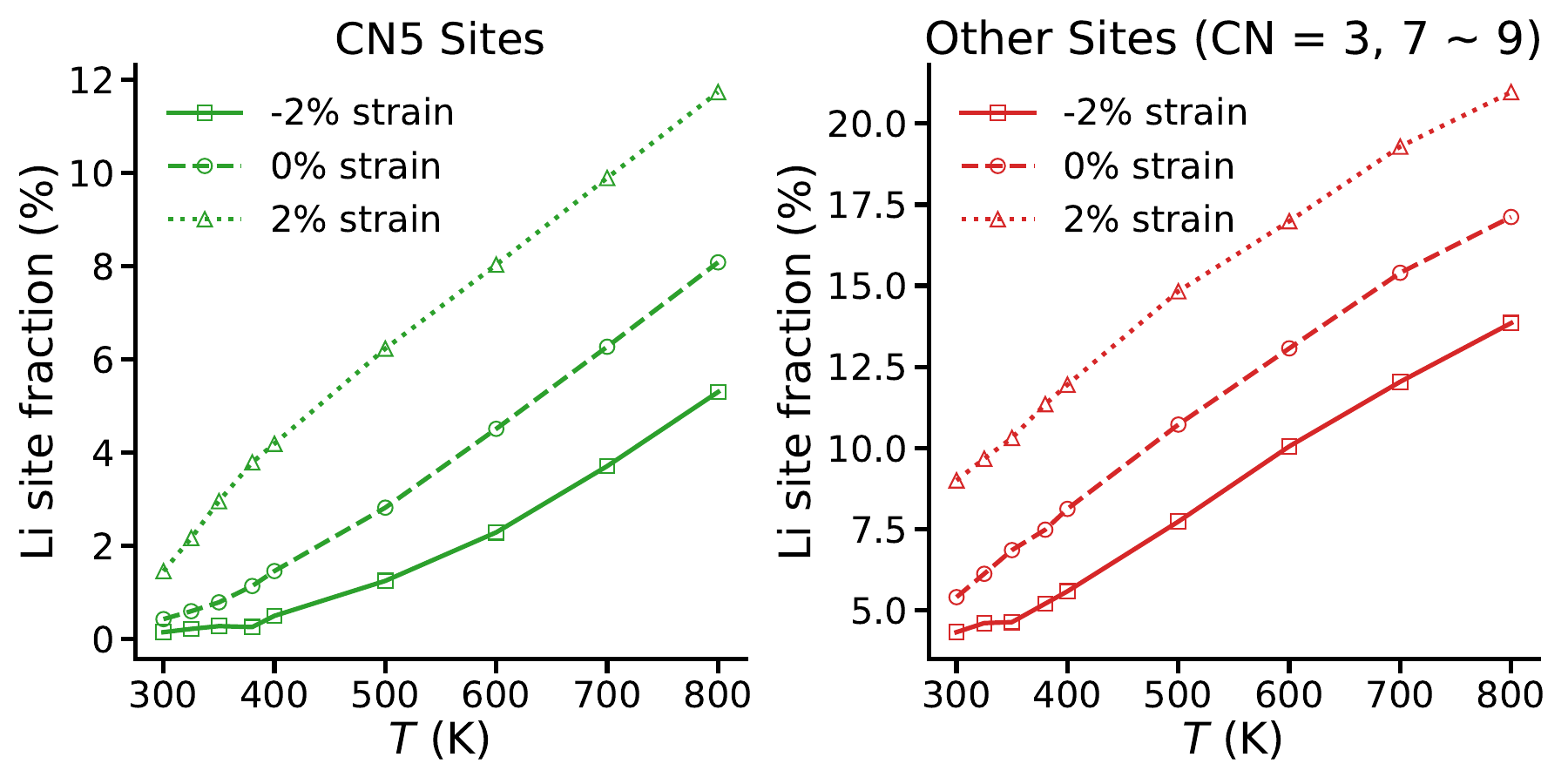}
  \caption{\textbf{\ce{Li+} site occupancy for the $CN = 5$ and other coordination numbers.}}
\end{figure}

\section{Elastic Moduli}
\label{sec:elastic:moduli}

Given the elastic tensor components $C_{ij}$ obtained from DFT calculations or interatomic potentials, the bulk modulus ($K$), shear modulus ($G$), and Young's modulus ($E$) can be computed as averaged measures for polycrystalline materials, reflecting the overall elastic response.
There are three common averaging schemes.
In the Voigt approach, uniform strain is assumed across all grains, yielding upper-bound estimates of the moduli.
In the Reuss approach, uniform stress is assumed, yielding lower-bound estimates.
The Hill approach takes the arithmetic mean of the Voigt and Reuss values and provides a more accurate representation~\cite{nye1985physical,wen2024equivariant}.
\begin{widetext}
  \begin{equation}
    \begin{aligned}
      K_V & = \frac{C_{11} + C_{22} + C_{33} + 2(C_{12} + C_{23} + C_{13})}{9},                                   \\
      G_V & = \frac{C_{11} + C_{22} + C_{33} - C_{12} - C_{23} - C_{13} + 3(C_{44} + C_{55} + C_{66})}{15}        \\
      K_R & = \frac{1}{S_{11} + S_{22} + S_{33} + 2(S_{12} + S_{23} + S_{13})},                                   \\
      G_R & = \frac{15}{4(S_{11} + S_{22} + S_{33}) - 4(S_{12} + S_{23} + S_{13}) + 3(S_{44} + S_{55} + S_{66})}, \\
      K_H & = \frac{K_V + K_R}{2}, \quad G_H = \frac{G_V + G_R}{2}.
    \end{aligned}
  \end{equation}
\end{widetext}
In all three schemes, the elastic moduli can be obtained as $E = 9KG/(3K + G)$ by using the corresponding $K$ and $G$ values.
Here, $S_{ij}$ denotes the components of the compliance tensor, obtained as the inverse of the elastic tensor: $\mathbf{S} = \mathbf{C}^{-1}$.
The reported bulk, shear, and Young's moduli in this work were calculated using the Hill averaging scheme.

\section{Stress Under Isotropic Strain}

For an isotropic strain of 2\%, the strain vector in Voigt notation is $\boldsymbol{\varepsilon} = (0.02,\,0.02,\,0.02,\,0,\,0,\,0)^\top$, and the stress follows from $\boldsymbol{\sigma} = \mathbf{C}\,\boldsymbol{\varepsilon}$.
Using the elastic tensor values from Table~II in the main text, we compute the stress.
For the \pmtmo phase,
\begin{equation}
  \begin{pmatrix}\sigma_1\\\sigma_2\\\sigma_3\\\sigma_4\\\sigma_5\\\sigma_6\end{pmatrix}
  =
  \begin{pmatrix}
    54.7 & 20.4 & 13.1 & -1.7 & 0    & 0    \\
    20.4 & 54.7 & 13.1 & 1.7  & 0    & 0    \\
    13.1 & 13.1 & 41.4 & 0    & 0    & 0    \\
    -1.7 & 1.7  & 0    & 13.7 & 0    & 0    \\
    0    & 0    & 0    & 0    & 13.7 & -1.7 \\
    0    & 0    & 0    & 0    & -1.7 & 17.2
  \end{pmatrix}
  \begin{pmatrix}0.02\\0.02\\0.02\\0\\0\\0\end{pmatrix}
  =
  \begin{pmatrix}1.764\\1.764\\1.352\\0\\0\\0\end{pmatrix}\ \text{GPa}.
\end{equation}
The maximum stress is $\sigma_1=\sigma_2=1.764$~GPa, attained in the basal plane, while the $c$-axis component $\sigma_3=1.352$~GPa is about 23\% smaller, reflecting the softer $C_{33}$.

For the \pnma phase,
\begin{equation}
  \begin{pmatrix}\sigma_1\\\sigma_2\\\sigma_3\\\sigma_4\\\sigma_5\\\sigma_6\end{pmatrix}
  =
  \begin{pmatrix}
    53.5 & 19.0 & 12.3 & 0    & 0    & 0    \\
    19.0 & 52.9 & 12.5 & 0    & 0    & 0    \\
    12.3 & 12.5 & 38.6 & 0    & 0    & 0    \\
    0    & 0    & 0    & 12.9 & 0    & 0    \\
    0    & 0    & 0    & 0    & 11.6 & 0    \\
    0    & 0    & 0    & 0    & 0    & 16.9
  \end{pmatrix}
  \begin{pmatrix}0.02\\0.02\\0.02\\0\\0\\0\end{pmatrix}
  =
  \begin{pmatrix}1.696\\1.688\\1.268\\0\\0\\0\end{pmatrix}\ \text{GPa}.
\end{equation}
The maximum stress is $\sigma_1=1.696$~GPa along the $a$ axis, narrowly exceeding $\sigma_2=1.688$~GPa along $b$, while $\sigma_3=1.268$~GPa along $c$ is again the smallest by roughly 25\%.

\section{\ce{Cl-} Vibrational Amplitude}
\label{sec:cl:amplitude}

To test whether the \ce{Cl-} framework is stiff at low temperature and looser on
heating, we measured the amplitude of the \ce{Cl-} vibrations.
Each \ce{Cl-} ion stays near one site, so we average its position over the NVT
trajectory at fixed cell to locate that site.
The amplitude $u$ is the typical distance between the ion and this site, taken as
the root mean square over time and over all \ce{Cl-} ions.
The amplitude stays between 0.2 and 0.5~\AA{} (\tref{tab:cl:amplitude}), a
small fraction of the 3.8~\AA{} Cl--Cl spacing, so the anions vibrate about
fixed sites rather than migrating.
The ratio $u/d$ grows from about 6\% at 300~K to about 11\% at 800~K in both
phases, so the anion framework does loosen appreciably over the range in which
the conductivity crossover occurs.

\begin{table}[tbh!]
  \centering
  \caption{\textbf{Vibrational amplitude of the \ce{Cl-} ions.}
    $u$ is the typical distance between each \ce{Cl-} ion and its own mean
    position, $d$ is the mean distance to the \ce{Cl-} neighbors in the first
    shell (within 5~\AA), and $u/d$ is their ratio.
    Trajectories are 1~ns, except at $-2$\% strain and 800~K, where they are
    480~ps (\pmtmo) and 570~ps (\pnma).}
  \label{tab:cl:amplitude}
  \begin{tabular}{ccrrrr}
    \hline
    Phase & Strain & $T$ (K) & $u$ (\AA) & $d$ (\AA) & $u/d$ (\%) \\
    \hline
    \pmtmo & 0\%    & 300 & 0.233 & 3.772 &  6.2 \\
           &        & 500 & 0.318 & 3.794 &  8.4 \\
           &        & 800 & 0.420 & 3.816 & 11.0 \\
           & $-2$\% & 800 & 0.413 & 3.797 & 10.9 \\
           & $+2$\% & 800 & 0.459 & 3.855 & 11.9 \\
    \pnma  & 0\%    & 300 & 0.245 & 3.778 &  6.5 \\
           &        & 500 & 0.325 & 3.799 &  8.6 \\
           &        & 800 & 0.431 & 3.822 & 11.3 \\
           & $-2$\% & 800 & 0.392 & 3.799 & 10.3 \\
           & $+2$\% & 800 & 0.471 & 3.864 & 12.2 \\
    \hline
  \end{tabular}
\end{table}

\clearpage

\def\bibsection{\section*{\refname}} 
\nocite{chen2024accelerating,park2020theoretical,qi2021bridging,schlem2021insights}
%